%% file: ntcir-arxiv.tex
\documentclass[graybox,natbib]{svmult}

\usepackage{mathptmx}       
\usepackage{helvet}         
\usepackage{courier}        
\usepackage{type1cm}        
\usepackage{makeidx}         
\usepackage{graphicx}        
\usepackage{multicol}        
\usepackage[bottom]{footmisc}

\usepackage{amsmath}
\usepackage{acronym}
\usepackage{url}

\let\cite\citep

\makeindex             

\begin{document}

\title*{Graded Relevance Assessments and\\ 
Graded Relevance Measures of NTCIR:\\ 
A Survey of the First Twenty Years}
\titlerunning{Graded Relevance Assessments and Graded Relevance Measures of NTCIR}
\author{Tetsuya Sakai
}
\institute{Tetsuya Sakai \at Waseda University, Shinjuku-ku Okubo 3-4-1 63-05-04,
Tokyo 169-8555, Japan \email{tetsuyasakai@acm.org}
}
%
%
\maketitle

\abstract*{
a
}

\abstract{
NTCIR was the first large-scale IR evaluation conference
to construct test collections with graded relevance assessments:
the NTCIR-1 test collections from 1998 already featured
\emph{relevant} and \emph{partially relevant} documents.
In this paper,
I first describe a few graded-relevance measures that originated from NTCIR (and a few variants)
which are used across different NTCIR tasks.
I then provide a survey on the use of graded relevance assessments
and of graded relevance measures in the past NTCIR tasks which primarily tackled ranked retrieval.
My survey shows that the majority of the past tasks fully utilised graded relevance 
by means of graded evaluation measures, but not all of them;
interestingly,
even a few relatively recent tasks chose to adhere to binary relevance measures.
I conclude this paper by a summary of my survey in table form,
and a brief discussion on what may lie beyond graded relevance.
}

\section{Introduction}\label{s:intro}

The evolution of NTCIR is quite different from that of TREC
when it comes to how \emph{relevance assessments}\index{relevance assessments}
have been conducted and utilised.
In 1992,
TREC started off with a high-recall task (i.e., the \emph{adhoc} track), with \emph{binary} relevance assessments~\cite[p.~34]{harman05}:
\begin{quotation}
Relevance was defined within the task of the information analyst,
with TREC assessors instructed to judge a document relevant
if information from that document would be used in some manner for the writing 
of a report on the subject of the topic.
This also implies the use of binary relevance judgments;
that is, a document either contains useful information and is therefore relevant,
or it does not.
\end{quotation}
Moreover, early TREC tracks heavily relied on evaluation measures based on binary 
relevance, such as 
\emph{11-point Average Precision},
\emph{R-precision}\index{R-precision}, and
(noninterpolated)
\emph{Average Precision}\index{average precision}~\cite{buckley05},
which meant,
for example, that
\emph{highly} relevant documents
and 
\emph{marginally} relevant documents~\cite{sormunen02}
were treated as if they were equally valuable.
It was in the TREC 2000 (a.k.a. TREC-9) Main Web task that 3-point \emph{graded} relevance assessments were introduced,
based on feedback from
web search engine companies at that time~\cite[p.~204]{hawking05}.
Accordingly, this task also adopted 
\ac{DCG},
proposed at SIGIR 2000~\cite{jarvelin00},
 to utilise the graded relevance assessments\footnote{
 A highly relevant document document 
 was considered to be worth 100 times as much as a relevant one.
 Natural logarithm was used as the patience parameter $b$ for discounting.
 }.

NTCIR has collected graded relevance assessments from the very beginning:
the NTCIR-1 test collections from 1998 already featured
\emph{relevant} and \emph{partially relevant} documents~\cite{kando99}.
Thus, 
while NTCIR borrowed many ideas from TREC when it was launched in the late 1990s,
its policy regarding relevance assessments 
seems to have
followed the paths of
\emph{Cranfield II}\index{Cranfield II} (which had 5-point relevance levels)~\cite[p.~21]{cleverdon66},
\ac{OHSUMED}\index{OHSUMED} (which had 3-point relevance levels)~\cite{hersh94},
as well as 
the first Japanese IR test collections \emph{BMIR-J1} and \emph{BMIR-J2} (which also had
3-point relevance levels)~\cite{sakai99bmir}.

Interestingly, with perhaps a notable exception of the aforementioned TREC 2000 Main Web Task,
it is true for both TREC and NTCIR that
the introduction of graded relevance assessments 
did not necessarily mean immediate adoption 
of \emph{evaluation measures} that can utilise graded relevance.
For example, although the TREC 2000 filtering track~\cite{robertson01}
reused the aforementioned OHSUMED collection, its evaluation measures were based on binary relevance;
while the TREC 2003-2005 robust tracks constructed adhoc IR test collections with 
3-point scale graded relevance assessments,
they adhered to binary-relevance measures such as \ac{AP}~\cite{voorhees06}.
Similarly, as I shall discuss in this paper,
while almost all of the past IR tasks of NTCIR had graded relevance assessments,
not all of them fully utilised them by means of graded-relevance measures.
This is the case despite the fact that
a graded-relevance measure called
the \emph{normalised sliding ratio} (NSR)
was proposed in 1968~\cite{pollock68},
and was discussed in an 1997 book by \citet{korfhage97},
along with another graded-relevance measure that is
based on user satisfaction and frustration.
NSR is actually what is now known
as \emph{normalised (nondiscounted) cumulative gain} (nCG)\footnote{
\citet[p.~209]{korfhage97} suggests that the ideal list (which Pollock calls the \emph{master list})
of NSR is obtained by reordering the top $k$ documents of a given system output,
and
\citet[p.426]{jarvelin02} argue that 
this is different from the ideal list for normalised (discounted) cumulative gain (n(D)CG).
However,
Korfhage's example is not accurate,
for
\citet{pollock68} defines his master list
as  ``\emph{a listing of all documents in the library [$\cdots$]
as being ordered in decreasing master value.}''
That is, his master list is exactly the ideal list of n(D)CG, 
and therefore his NSR is exactly nCG.
However, this measure is a set retrieval measure;
unlike nDCG, it is not adequate as a ranked retrieval measure.
See Section~\ref{sss:Q} for the formal definition of NSR / nCG.
}.

Section~\ref{s:measures} briefly describes a few graded-relevance
measures that originated from NTCIR and have been used in several NTCIR tasks.
Section~\ref{s:binary} provides an overview of 
past ranked retrieval tasks of NTCIR that adhered to binary-relevance measures
despite having graded relevance assessments.
Section~\ref{s:graded} provides and overview 
of past ranked retrieval tasks of NTCIR that utilised graded-relevance measures.
Finally, 
Section~\ref{s:summary} summarises the above survey in table form,
and discusses what may lie beyond evaluation based on graded relevance.

It should be noted that the present survey covers only NTCIR tasks 
that are primarily ranked retrieval and involve graded relevance assessments:
Primarily
NLP-oriented tasks such as summarisation and question answering
are outside the scope; also, \emph{Crosslingual Link Discovery}~\cite{tang13}
is not discussed here as the task did not have any graded relevance assessments
although it did involve ranked retrieval.


\section{A Few Graded Relevance Measures Used across NTCIR Tasks}\label{s:measures}

\subsection{Q-measure for Adhoc IR, and its Variants}\label{ss:Q}

This section briefly describes the Q-measure (or just ``Q'' for short)
and its variants. All of them are graded-relevance measures
for ad hoc IR.
The measures discussed in this section 
can be computed 
using \texttt{NTCIREVAL}\footnote{
\url{http://research.nii.ac.jp/ntcir/tools/ntcireval-en.html}
}
or its predecessor \texttt{ir4qa\_eval}~\cite{sakai08ir4qa}\footnote{
\url{http://research.nii.ac.jp/ntcir/tools/ir4qa_eval-en.html}
}.

\subsubsection{Q-measure}\label{sss:Q}

The Q-measure bears its name because it was originally
designed for evaluating a ranked list of answers 
for a given question, where 
multiple correct answer strings could 
form an equivalence class~\cite{sakai04ntcir}.
For example,
for a question ``{\it Who were the members of The Beatles?},''
a gold equivalence class 
could contain
``{\it Ringo Starr}'' (highly relevant) and
``{\it Richard Starkey}'' (partially relevant).
If both of the above answer strings 
are included in the same ranked list of answers,
then only one of them is treated as relevant\footnote{
A similar idea was used earlier in the NTCIR-3 Web Retrieval task~\cite{eguchi03},
where retrieving duplicate web pages was penalised (See Section~\ref{ss:web}).
}.
However, for document retrieval
where we have document IDs instead of arbitrary answer strings,
the notion of equivalence class disappears.
In this situation, Q 
is actually a generalised form of \ac{AP},
as explained below.

Given a graded-relevance test collection,
we follow the approach of nDCG
and first decide on the \emph{gain value} ${\it gv}_{x}$ for 
each relevance level $x$;
for $x=0$ (nonrelevant), we let ${\it gv}_{0}=0$.
For a given ranked list of documents for a particular topic,
we let the gain at rank $r$ be $g(r)={\it gv}_{x}$ 
if the document at $r$ is $x$-relevant.
Moreover, for this topic,
we consider an ideal ranked list,
obtained by listing up all
known $x$-relevant ($x>0$)
documents in decreasing order of $x$;
we denote the gain at rank $r$ in this ideal list by $g^{\ast}(r)$.
Let the \emph{cumulative gain} at rank $r$ of the system output
be ${\it cg}(r)=\sum_{i=1}^{r}g(i)$; similarly, let ${\it cg}^{\ast}(r)=\sum_{i=1}^{r}g^{\ast}(i)$.
Note that nCG (i.e., NSR) at cutoff $l$
is exactly ${\it cg}(l)/{\it cg}^{\ast}(l)$.
However, this measure is not adequate as a ranked retrieval measure,
since the ranks of the retrieved relevant documents within top $l$
do not matter.

Let the total number of relevant documents (i.e., $x$-relevant where $x>0$)
for the topic be $R$.
Let $I(r)$ be a flag, which equals zero if the document at $r$ is nonrelevant
and one otherwise. Then $C(r)=\sum_{i=1}^{r}I(i)$ is the number of relevant 
documents between ranks 1-$r$. The Q-measure is defined as
\begin{equation}\label{eq:Q}
Q = \frac{1}{R} \sum_{r} I(r) {\it BR}(r) \ ,
\end{equation}
where ${\it BR}(r)$ is the \emph{blended ratio}\index{blended ratio}
given by
\begin{equation}\label{eq:BR}
{\it BR}(r) = \frac{C(r) + \beta {\it cg}(r)}{r + \beta {\it cg}^{\ast}(r)} \ .
\end{equation}
Here, $\beta$ is the \emph{patience parameter} which is usually set to one;
its significance is discussed in \citet{sakai14promise}.
Note that $C(r)/r$ represents binary \emph{Precision} at rank $r$;
hence, both precision and nCG are embedded in Eq.~\ref{eq:BR}.
Moreover, note that letting $\beta=0$ reduces Q to \ac{AP}.

Just as nDCG penalises relevant documents retrieved at low ranks 
by means of discounting the gain values,
Q achieves a similar effect by means of the $r$ in the denominator of Eq.~\ref{eq:BR};
see \citet{sakai14promise} for details.
Q and nDCG behave quite similarly; see, for example, \citet{sakai06sigir} and~\citet{sakai07ipm}.
For a comparison of 
Q and \emph{Graded Average Precision}~\cite{robertson10}, see \citet{sakai11sigir};
for a comparison of Q and \emph{Generalised Average Precision}~\cite{kishida05genap},
see \citet{sakai07evia}.

For small document cutoffs (e.g., evaluating the top 10 URLs in a \ac{SERP}),
the following cutoff-based Q may also be used, to ensure the $[0,1]$ range:
\begin{equation}\label{eq:Qcutoff}
Q@l = \frac{1}{\min(l,R)} \sum_{r}^{l} I(r) {\it BR}(r) \ ,
\end{equation}
where $l$ is the cutoff value\footnote{
Note that if $R>l$, Q as defined by Eq.~(\ref{eq:Q}) would be smaller than 1 even for an ideal list.
}.

As we shall discuss later,
Q was used in the following NTCIR ranked retrieval tasks:
CLIR~\cite{kishida07},
IR4QA~\cite{sakai08ir4qa,sakai10ir4qa},
GeoTime~\cite{gey10,gey11},
CQA~\cite{sakai10cqa},
Temporalia~\cite{joho14},
WWW~\cite{luo17,mao19},
OpenLiveQ~\cite{kato17,kato19},
and 
CENTRE~\cite{sakai19centre}.

\subsubsection{Variants}\label{sss:Qvariants}

Just like \ac{AP} and nDCG,
Q is more suitable for topics with \emph{informational} intents
than \emph{navigational} ones~\cite{broder02}:
retrieving many relevant documents is rewarded.
On the other hand, \emph{P$^{+}$} is a variant of Q
suitable for topics with navigational intents, for which
just one or a few highly relevant documents are needed.
As I shall discuss in Section~\ref{ss:STC},
P$^{+}$ was used in the NTCIR-12 and -13 \ac{STC} tasks~\cite{shang16,shang17}.

For a given system output and a document cutoff $l$, let $r_{p}$ 
be the rank of the highest-ranked document among the 
\emph{most relevant} documents within top $l$.
This is called the \emph{preferred rank},
denoted by $r_{p}$.
The model behind 
\emph{P$^{+}$} assumes that 
no user will examine documents below rank $r_{p}$.
If there is no relevant document within top $l$, then P$^{+}=0$.
Otherwise,
\begin{equation}\label{eq:Pplus}
P^{+} = \frac{1}{C(r_{p})} \sum_{r}^{r_{p}} I(r) {\it BR}(r) \ .
\end{equation}
For a comparison of P$^{+}$ with
other measures designed for evaluation with navigational intents 
such as \ac{WRR}~\cite{eguchi03},
see \citet{sakai07tod}. Section~\ref{ss:web} also touches upon \ac{WRR}.

Both Q and P$+$ are instances of the \ac{NCU} family~\cite{sakai08evia},
defined as 
\begin{equation}\label{eq:NCU}
{\it NCU} = \sum_{r=1}^{\infty} P_{S}(r) {\it NU}(r) \ ,
\end{equation}
Given a population of users, it is assumed that 
$100P_{S}(r)$\% of those users will stop and abandon the ranked list at rank~$r$ due to satisfaction;
the utility of the ranked list for this particular group of users is given by ${\it NU}(r)$.
Hence, NCU is the expectation of the normalised utility over a population of users.
For Q, the stopping probability $P_{S}(r)$ is uniform over all relevant documents;
for P$+$, it is uniform over all relevant documents at or above rank $r_{p}$.
Both Q and P$+$ use ${\it BR}(r)$ as the utility.
Chapelle {\it et al.}~\cite{chapelle09} point out that their {\em Expected Reciprocal Rank} (ERR)
measure is also an instance of NCU:
ERR's stopping probability\footnote{
\citet{sakai08evia} discuss a stopping probability distribution 
that also depends on relevant documents seen so far, which they call the rank-biased distribution.
} at $r$ depends on the relevant documents seen within ranks 1-$(r-1)$,
and its utility is given by the \ac{RR}: $1/r$.
Precision at $l$, \ac{AP}, and \ac{RR} can also be regarded as instances of NCU\footnote{
Precision at $l$ assumes that all users stop at rank $l$;
\ac{AP} assumes that the stopping probability distribution is uniform over all relevant documents;
RR assumes that all users stop at rank $r_{1}$, the rank of the first relevant document in the SERP.
For all these measures, the utility at $r$ is given by Precision at $r$.
}.

\subsection{D$\sharp$-measures for Diversified IR, and its Variants}

This section briefly describes the D$\sharp$-measures~\cite{sakai11sigir}
and their variants. All of them are
graded-relevance measures 
for diversified IR.
While the diversified IR tasks at TREC
have used $\alpha$-nDCG~\cite{clarke08}
and intent-aware measures~\cite{agrawal09}
(e.g.
intent-aware \ac{ERR}~\cite{chapelle11}) etc.,
the diversified IR tasks of NTCIR
have used D$\sharp$-measures
and their variants as the official measures.

For diversified IR evaluation,
we generally require a set of topics,
and a set of \emph{intents} for each topic,
and the probability of intent $i$ given topic $q$ (${\it Pr}(i|q)$),
and \emph{intentwise} graded relevance assessments.
Note that
an adhoc IR test collection may be regarded as 
a specialised case of a diversified IR test collection where
every topic has exactly one intent $i$ such that ${\it Pr}(i|q)=1$.

The measures discussed in this section 
can also be computed 
using \texttt{NTCIREVAL} (See Section~\ref{ss:Q}).

\subsubsection{D$\sharp$-measure}\label{sss:D}

A D$\sharp$-measure (e.g., D$\sharp$-nDCG)
is defined as:
\begin{equation}\label{eq:Dsharp}
D\sharp\mbox{-}measure = \gamma I\mbox{-}rec + (1-\gamma) D\mbox{-}measure \ ,
\end{equation}
where I-rec is the \emph{intent recall} (a.k.a., \emph{subtopic recall}~\cite{zhai03}),
i.e., the number of intents covered by the \ac{SERP} for a given topic;
$\gamma$ is a parameter that balances I-rec (a pure diversity measure)
and D-measure (an overall relevance measure explained below), usually set to $\gamma=0.5$.

A D-measure (e.g., D-nDCG)
is a measure defined by first constructing an ideal ranked list
based on the \emph{global gain} for each document
and then computing an adhoc IR measure (e.g., nDCG)
based on the ideal list. 
For each document, if it is $x$-relevant to intent $i$,
we give it a gain value of ${\it gv}_{i,x}$;
its global gain is then computed as $\sum_{i} {\it Pr}(i|q) {\it gv}_{i,x}$.
A global ideal list is then obtained by sorting all documents by the global gain (in descending order).
Thus, unlike intent-aware measures~\cite{agrawal09,chapelle11},
a \emph{single} ideal list is defined for a given topic $q$.
Similarly,
the system's ranked list is also scored using the global gain:
if the document at $r$ is $x$-relevant to intent $i$ (for each $i$),
we let the intentwise gain value be $g_{i}(r)={\it gv}_{i,x}$,
so that 
the global gain for this document is given by
\begin{equation}\label{eq:GG}
{\it GG}(r)=\sum_{i} {\it Pr}(i|q) g_{i}(r) \ .
\end{equation}
Thus graded-relevance measures such as nDCG and Q can be computed
by treating the global gain values just like traditional gain values.
The resultant measures are called D-nDCG, D-Q, etc.

The principle behind the global ideal list is as follows.
Let ${\it rel}$ be a random binary variable: it can either be 1 (relevant) or 0 (nonrelevant).
In diversity evaluation where we have a set of intents $\{i\}$ for topic $q$,
we define ${\it rel}$ to be 1 for $(q,d)$ 
iff there is at least one intent $i$ for $q$ such that $d$ is relevant for $i$.
Given a topic $q$ and a set of documents $\{d\}$, 
the \ac{PRP} dictates that we rank the documents by ${\it Pr}({\it rel}=1|q,d)$,
that is, the probability that $d$ is relevant to $q$.
If we assume that the intents are mutually exclusive,
the above can be rewritten as $\sum_{i}{\it Pr}(i|q) {\it Pr}({\it rel}=1|i,d)$,
where ${\it Pr}({\it rel}=1|i,d)$ is the probability that $d$ is relevant to $i$.
If we further assume that ${\it Pr}({\it rel}=1|i,d)$ is proportional to the 
intentwise gain value ${\it gv}_{i,x}$, 
the sort key for defining the global ideal list becomes $\sum_{i} {\it Pr}(i|q) {\it gv}_{i,x}$,
which is exactly the global gain.

As we shall discuss later, D$\sharp$-measures have been used 
in the the following NTCIR ranked retrieval tasks:
INTENT~\cite{song11,sakai13intent}
IMine~\cite{liu14,yamamoto16}, and
Temporalia~\cite{joho16}.
For comparisons of D$\sharp$-measures with 
\emph{intent-aware} measures~\cite{agrawal09,chapelle11} and $\alpha$-nDCG,
we refer the reader to \citet{sakai11sigir,sakai13irj}.
Other studies that compared these different types of diversity measures 
or their extensions
include \citet{golbus13,zhou13,amigo18}.
D$\sharp$-measures have also been extended to handle
\emph{hierarchical} intents~\cite{hu15,wang18}; see also Section~\ref{sss:Dvariants}.

\subsubsection{Variants Used at NTCIR}\label{sss:Dvariants}

In the NTCIR-10 INTENT-2 task~\cite{sakai13intent},
each intent for a topic 
had not only the intent probability ${\it Pr}(i|q)$,
but also a tag indicating whether it is \emph{informational} or \emph{navigational}.
For informational intents, retrieving more relevant documents should be rewarded;
for naviational intents, this may not be a good idea.
Hence, the task employed a \emph{DIN-measure}\footnote{DIN stands
for diversification with informational and navigational intents}~\cite{sakai12www}
in addition to D$\sharp$-measures.
The only difference between D-measures and DIN-measures lies in 
how the global gain is computed for each document in the system output;
the ideal list is unchanged.
Let $\{i\}$ and $\{j\}$ denote the sets of 
informational and navigational intents for topic $q$,
and let ${\it isnew}_{j}(r)=1$ if there is no document relevant to 
the navigational intent $j$ between ranks 1 and $r-1$,
and ${\it isnew}_{j}(r)=0$ otherwise.
Then, for DIN-measures, the global gain for the document at rank $r$ in the system output is given by
\begin{equation}\label{eq:DIN}
GG^{\it DIN}(r) = \sum_{i} {\it Pr}(i|q)g_{i}(r) + \sum_{j} {\it isnew}_{j}(r) {\it Pr}(j|q) g_{j}(r) \ .
\end{equation}
Thus, ``redundant'' relevant documents for each navigational intent are ignored.
The instance of DIN-measure actually used at the task was DIN-nDCG.

Also at the NTCIR-10 INTENT-2 task,
an extension of the Q-measure and P$+$
was used as an additional measure to
handle the informational and navigational intents.
For each informational intent $i$,
let $Q_{i}@l$ be the score computed using Eq.~\ref{eq:Qcutoff}
by treating only documents relevant to $i$ as relevant (i.e., intentwise Q);
for each navigational intent $j$,
 let $P_{j}$ be the score computed using Eq.~\ref{eq:Pplus}
by treating only documents relevant to $j$ as relevant (i.e., intentwise P$+$).
The overall diverisity measure, called P$+$Q~\cite{sakai12www}, is given by
\begin{equation}
P+Q@l = \sum_{i} {\it Pr}(i|q)Q_{i}@l + \sum_{j} {\it Pr}(j|q)P_{j}^{+} \ .
\end{equation}
Note that P+Q is a type of intent-aware measure; the novelty is 
that different measures (or actually, different stopping probability distributions) are used across different intents.

The Subtopic Mining (SM) subtask of 
the NTCIR-11 IMine Task~\cite{liu14}
required systems to return three files for a given query:
(a) a two-level hierarchy of subtopics;
(b) a ranked list of the first-level subtopics; and
(c) a ranked list of the second-level subtopics,
in contrast to the SM subtask of INTENT task, which only required Output~(b).
Accordingly, the IMine organisers introduced the \emph{H-measure} for the subtask:
\begin{equation}
H\mbox{-}measure = {\it Hscore} (\alpha D^{1}\sharp\mbox{-}measure + (1-\alpha) D^{2}\sharp\mbox{-}measure) \ ,
\end{equation}
where $D^{1}\sharp\mbox{-}measure$ is the D$\sharp$-measure score computed 
for Output~(b), $D^{2}\sharp\mbox{-}measure$ is the D$\sharp$-measure score computed 
for Output~(c),
and ${\it Hscore}$ is the fraction of second-level subtopics that are correctly assigned to their
first-level subtopics, i.e., the accuracy of Output~(a).
The NTCIR-11 IMine Task actually computed D$\sharp$-nDCG as an instance of the D$\sharp$-measure family.

The Query Understanding (QU) subtask of the NTCIR-12 IMine-2 Task~\cite{yamamoto16}
was similar to the previous SM subtasks,
but requires systems 
to return a ranked list of $($subtopic, vertical$)$ pairs (e.g., $($``\textit{iPhone 6 photo}'', Image$)$, 
$($``\textit{iPhone 6 review}'', Web))
for a given query.
Accordingly, they used the following extension of the D$\sharp$-nDCG:
\begin{equation}\label{eq:QU}
QU\mbox{score}@l = \lambda D\sharp\mbox{-}nDCG@l + (1-\lambda)V\mbox{-}score@l \ ,
\end{equation}
where $V\mbox{-}score$ measures 
the appropriateness of the named vertical for each subtopic in the system list
by leveraging ${\it Pr}(v|i)$, i.e.,
 the importance of vertical $v$ for intent $i$.
More specifically, it is given by
\begin{equation}\label{eq:Vscore}
V\mbox{-}score@l = \frac{1}{l} \sum_{r=1}^{l} \frac{{\it Pr}(v(r)|i(r))}{\max_{v \in V} {\it Pr}(v|i(r))} \ ,
\end{equation}
where $V$ is the set of verticals,
$v(r)$  is the vertical returned by the system at rank $r$,
and $i(r)$ is the intent to which the subtopic returned at rank $r$ belongs\footnote{
In INTENT and IMine tasks,
an \emph{intent} is derived by manually clustering a set of submitted \emph{subtopic} strings.
Therefore, a subtopic belongs to exactly one intent (See also Section~\ref{ss:intent}).}.

\section{Graded Relevance Assessments, Binary Relevance Measures}\label{s:binary}

This section provides an overview of NTCIR ranked retrieval tasks that did not 
use graded-relevance evaluation measures even though they had graded relevance assessments.

\subsection{Early IR and CLIR Tasks (NTCIR-1 through -5)}\label{ss:earlyIR}

The Japanese IR and (Japanese-English) crosslingual tasks of NTCIR-1~\cite{kando99}
constructed test collections with
3-point relevance levels (relevant, partically relevant, nonrelevant),
but used binary-relevance measures such as \ac{AP} and \emph{R-precision}\footnote{
For a topic with $R$ known relevant documents,
R-precision is the precision at rank $R$, or equivalently, the recall at rank $R$.
Note that this is a set retrieval measure, not a ranked retrieval one.
} by either treating the relevant and partially relevant documents as ``relevant'' 
or treating only the relevant documents as ``relevant.''
However, 
it should be stressed at this point that using binary-relevance measures with
different relevance thresholds cannot serve as substitutes
for a graded relevance measure that encourages optimisation towards 
an ideal ranked list (i.e., a list of documents sorted in decreasing order of relevance levels).
If partially relevant documents are ingnored, 
a SERP whose top $l$ documents are all partially relevant
and one whose top $l$ documents are all nonrelevant
can never be distinguished from each other;
if relevant documents and partially relevant documents are all treated as relevant,
a SERP whose top $l$ documents are all relevant
and one whose top $l$ documents are all partially relevant
can never be distinguished from each other.

The Japanese and English (monolingual and crosslingual) IR tasks of NTCIR-2~\cite{kando01}
constructed test collections with 4-point relevance levels: 
S (highly relevant), 
A (relevant), 
B (partially relevant),
and 
C (nonrelevant).
However, the organisers used binary-relevance measures such as AP and R-precision
by either treating only the S and A documents as relevant
or treating the S, A, and B documents as relevant.
As for the Chinese monolingual and Chinese-English IR tasks of NTCIR-2~\cite{chen01},
three judges independently judged each pooled document
using 4-point relevance levels, and then a score was assigned to 
each relevace level. Finally, the scores were average across the three assessors.
For example, if a document is judged ``very relevant'' (score: 1),
``relevant'' (score: 2/3), and ``partially relevant'' (score: 1/3) independently,
its overall score is $(1+2/3+1/3)/3=2/3$.
The organisers then applied two different thresholds
to map the scores to binary relevance:
\emph{rigid relevance} was defined 
by treating documents with scores $2/3$ or larger as relevant;
\emph{relaxed relevance} was defined 
by treating those with scores $1/3$ or larger as relevant.
For evaluating the runs,
rigid and relaxed versions of recall-precision curves (RP curves) were used.

The NTCIR-3 CLIR (Cross-Language IR) task~\cite{chen02} was similar to the previous IR tasks:
4-point relevance levels (S,A,B,C) were used;
rigid relevance was defined using the S and A documents
while relaxed relevance was defined using the B documents in addition;
finally, rigid and relaxed versions of AP were computed for each run.
The NTCIR-4 and NTCIR-5 CLIR tasks~\cite{kishida04,kishida05} adhered to the above practice.

It is worth noting that all of the above tasks used the {\tt trec\_eval} programme from TREC\footnote{
\url{https://trec.nist.gov/trec_eval/}
}
to compute binary-relevance measures such as AP.
At NTCIR-6, the CLIR organisers finally took up graded-relevance measures,
as I shall discuss in Section~\ref{ss:clir6}.

\subsection{Patent (NTCIR-3 through -6)}\label{ss:patent}

The Patent Retrieval Tasks of NTCIR,
which were run from NTCIR-3 (2002) to NTCIR-6 (2007),
never used graded-relevance measures 
despite having graded relevance assessments with highly unique properties.

The NTCIR-3 Patent Retrieval task~\cite{iwayama03} was
a news-to-patent \emph{technical survey} search task,
with 4-point relevance levels:
A (relevant), 
B (partially relevant),
C (judged nonrelevant after reading the entire patent), and
D (judged nonrelevant after just reading the patent title).
RP curves were 
drawn based on \emph{strict relevance} (only A treated as relevant)
and \emph{relaxed relevance} (A and B treated as relevant).
In the overview paper,
``median of the average precisions for each topic''
is discussed, but systems were not compared based on AP.

The main task of the NTCIR-4 Patent Retrieval task~\cite{fujii04}
was a patent-to-patent \emph{invalidity search} task.
There were two types of relevant documents:
\begin{description}
\item[A] A patent that can invalidate a given claim on its own;
\item[B] A patent that can invalidate a given claim
\emph{only when used with one ore more other patents}.
\end{description}
For example, patents $B_1$ and $B_2$ may each be nonrelevant
(as they cannot invalidate a claim individually),
but if they are both retrieved, the pair should serve as one relevant document.
In addition, the organisers provided passage-level \emph{binary} relevance 
assessments: if a \emph{single} passage provides sufficient grounds 
for the patent (from which the passage was drawn) to be either A or B, that passage is relevant;
if a \emph{group} of passages serves the same purpose,
that passage group is relevant.
However, these passage-level relevance assessments were not utilised for evalution at NTCIR-4.
At the evaluation step, 
the organisers used AP
by either treating only the A documents as relevant (rigid evaluation)
or treating both A and B documents as relevant (relaxed evaluation).
Note that the above relaxed evaluation 
has a limitation: recall the aforementioned example 
with $B_1$ and $B_2$, and consider 
a SERP that managed to return only one of them (say $B_1$).
Relaxed evaluation rewards the system for returning $B_1$,
even though this document alone does \emph{not} invalidate the claim.

The Document Retrieval subtask of the NTCIR-5 Patent Retrieval task~\cite{fujii05}
was similar to its predecessor, but the relevant documents were determined
purely based on whether and how they were actually
used by a patent examiner to reject a patent application;
no manual relevance assessments were conducted for this subtask.
The graded relevance levels were defined as follows:
\begin{description}
\item[A] A citation that was actually used on its own to reject a given patent application;
\item[B] A citation that was actually used along with another one to reject a given patent application.
\end{description}
As for the evaluation measure for Document Ranking,
the organisers adhered to rigid and relaxed AP.
In addition, the task organisers introduced a Passage Retrieval subtask
by leveraging
passage-level binary relevance assessments
collected as in the NTCIR-4 Patent task:
given a patent, systems were required to rank the passages from that same patent.
As both single passages and groups of passages 
can potentially be relevant to the source patent (i.e., the passage(s) 
can serve as evidence to determine that the entire patent is relevant to a given claim),
this poses a problem similar to the one discussed above with patents $B_1$ and $B_2$:
for example, if two passages $p_1, p_2$ are relevant as a group but not individually,
and if $p_1$ is ranked at $i$ and $p_2$ is ranked at $i' (>i)$,
how should the SERP of passage be evaluated?
To address this, the task organisers
introduced a binary-relevance 
measure called the \emph{Combinational Relevance Score} (CRS),
which assumes that the user who scans the SERP must reach
as far as $i'$ to view both $p_1$ and $p_2$\footnote{
In fact, AP, Q 
or any measure from the NCU family (See Section~\ref{sss:Qvariants})
can easily be extended to handle 
\emph{combinational relevance}
for Document Retrieval (See the above example with $(B_1, B_2)$)
and for Passage Retrieval (See the above example with $(p_1, p_2)$)~\cite{sakai06ipsj}.
For example, given a SERP that contains $B_1$ at rank~$i$ and $B_2$ at rank $i' (>i)$,
we can assume that a group of users 
will abandon the ranked list at rank $i'$, that is, only after viewing both documents.
Hence, for this user group, the utility (i.e., precision in the case of AP)
can be computed at 
rank $i'$,
but not at rank $i$.
}.

The Japanese Document Retrieval subtask of the NTCIR-6 Patent Retrieval task~\cite{fujii07}
had two different sets of graded relevance assessments;
the first set (``Def0'' with A and B documents) was defined in the same way as in NTCIR-5;
the second set (``Def1'') was automatically derived from Def0 based on the \ac{IPC} codes
as follows:
\begin{description}
\item[H] The set of IPC subclasses for this cited patent has no overlap with that of the input patent (and therefore it is relatively difficult to retrieve this patent);
\item[A] The set of IPC subclasses for this cited patent has some overlap with that of the input patent;
\item[B] The set of IPC subclasses for this cited patent is identical to that of the input patent (and therefore it is relatively easy to retrieve this patent).
\end{description}
As for the English Document Retrieval subtask,
the relevance levels were also automatically determined based on IPC codes,
but only two types of relevant documents were identified,
as each USPTO patent is given only one IPC code:
A (cited patents whose IPC subclasses were not identical to those of the input patent), and
B (cited patents whose IPC subclasses were identical to those of the input patent).
In both subtasks,
AP was computed by considering different combinations of the above relevance levels.

\subsection{SpokenDoc/SpokenQuery\&Doc (NTCIR-9 through -12)}

The NTCIR-9 and NTCIR-10 SpokenDoc tasks (2011, 2013)~\cite{akiba11,akiba13}
and the NTCIR-11 and NTCIR-12 SpokenDoc\&Query tasks (2014, 2016)~\cite{akiba14,akiba16}
also never used graded-relevance measures 
despite having graded relevance assessments.
Hereafter, we omit  the discussion of the Spoken Term Detection (STD) subtasks of 
these tasks as they did not involve graded relevance assessments.

The Spoken Document Retrieval (SDR) subtask of the NTCIR-9 SpokenDoc task
had two ``subsubtasks'': \emph{Lecture Retrieval}
and \emph{Passage Retrieval}, where a passage is any sequence of consecutive inter-pausal units.
Passage-level relevance assessments were obtained on a 3-point scale (relevant, partially relevant, and nonrelevant),
and it appears that the document-level (binary) relevance was deduced from them\footnote{
The official test collection data of the NTCIR-9 SDR task (\texttt{evalsdr}) contains
only passage-level gold data.
}.
AP was used for evaluating Document Retrieval, whereas
variants of AP, called \emph{utterance-based} (M)AP,
\emph{pointwise} (M)AP,
and
\emph{fractional} (M)AP
were used for evaluating Passage Retrieval.
While these variants compare the system's ranked list of passages
with the gold list of passages in different ways,
none of them utilise the distinction between
relevant and partially relevant passages in the gold data;
they are binary-relevance measures\footnote{
While the fractional (M)AP considers 
the degree of overlap between 
a gold passage and a retrieved passage,
whether the gold passage is relevant
or partially relevant is not considered.
}.
The NTCIR-10 SpokenDoc-2 Spoken Content Retrieval (SCR) subtask~\cite{akiba13}
was similar to the SDR subtask at NTCIR-9,
with Lecture Retrieval and Passage Retrieval subsubtasks.
Lecture Retrieval 
used a revised version of the NTCIR-9 SpokenDoc topic set,
and its gold data 
does not contain graded relevance assessments\footnote{
This was verified by examining 
\url{SpokenDoc2-formalrun-SCR-LECTURE-golden_20130129.xml}
in the SpokenDoc-2 test collection.
}; binary-relevance AP was used for the evaluation.
As for Passage Retrieval, a new topic set was devised,
again with 3-point relevance levels\footnote{
This was verified by examining
\url{
SpokenDoc2-formalrun-SCR-PASSAGE-golden_20130215.xml}
in the SpokenDoc-2 test collection.
}. The three binary-relevance AP variants from the NTCIR-9 SDR task
the evaluation was done in the same way as in NTCIR-9.

The Slide Group Segment (SGS) Retrieval subsubtask of the 
NTCIR-11 SpokenQuery\&Doc SCR subtask
involved the ranking of predefined retrieval units (i.e., SGSs),
unlike the Passage Retrieval subsubtask
which allows any sequence of consecutive inter-pausal units as a retrieval unit.
Three-point relevance levels were used to judge 
the SGSs:
R (relevant), P (partially relevant), and I (nonrelevant).
However, the binary AP was used for the evaluation.
As for the passage-level relevance assessments,
they were derived from the SGSs labelled R or P,
and were considered binary;
the three binary-relevance AP variants were used for this subsubtask once again.
Segment Retrieval was continued at the NTCIR-12 SpokenQuery\&Doc-2 task,
again with the same 3-point relevance levels and AP as the evaluation measure\footnote{
The NTCIR-12 SpokenQuery\&Doc-2 overview paper does not discuss
the evaluation of
Passage Retrieval runs.
}.

\subsection{Math/MathIR (NTCIR-10 through -12)}\label{ss:math}

The NTCIR-10 and NTCIR-11 Math tasks (2013, 2014)~\cite{aizawa13,aizawa14}
and the NTCIR-12 MathIR task (2016)~\cite{zanibbi16}
also adhered to binary-relevance measures despite having graded relevance assessments.

In the Math Retrieval subtask of the NTCIR-10 Math Task,
retrieved mathematical formulae were 
judged on a 3-point scale (relevant, partially relevant, nonrelevant).
Up to two assessors judged each formula,
and initially 5-point relevance scores were devised
 based on the results.
 For example, for formulae judged by one assessor,
 they were given 4 points if the judged label was relevant;
 for those judged by two assessors,
 they were given 4 points if both of them gave them the relevant label.
 Finally, the scores were mapped to a 3-point scale:
 Documents with scores 4 or 3 were treated as elevant;
 those with 2 or 1 were treated as partially relevant;
 those with 0 were treated as ronrelevant.
 However, at the evaluation step,
 only binary-relevance measures such as 
 AP and Precision were computed
 using \texttt{trec\_eval}.
 Similarly, in the Math Retrieval subtask of the NTCIR-11 Math Task,
 two assessors independently judged each retrieved unit 
 on a 3-point scale, 
 and the final relevance levels were also on a 3-point scale.
 If the two assessor labels 
 were relevant/relevant
 or relevant/partially-relevant, 
 the final grade was relevant;
 if the two labels were both nonrelevant,
 the final grade was nonrelevant;
 the other combinations were considered partially relevant.
 As for the evaluation measures,
 \emph{bpref}~\cite{buckley04,sakai07sigir,sakai08irj} was computed 
 along with AP and Precision, using \texttt{trec\_eval}.
 
 The NTCIR-12 MathIR task was similar to the Math Retrieval subtask of the 
 aforementioned Math tasks.
 Up to four assessors judged each retrieved unit
 using a 3-point scale,
 and the individual labels were consolidated to form
 the final 3-point scale assessments.
 As for the evaluation, only Precision (computed at several cutoffs) was used using \texttt{trec\_eval}.
 
 The following remarks from the Math and MathIR overview papers may be noteworthy:
 \begin{quotation}
 Since trec\_eval only accepts binary relevance judgment, the scores of the two judges were converted into an overall relevance score [$\ldots$]~\cite{aizawa14}
 \end{quotation}
 \begin{quotation}
 Since the trec\_eval tool only accepts binary relevance
judgments, the scores of evaluators were first converted into
a combined relevance score [$\ldots$]~\cite{zanibbi16}
  \end{quotation}
 This suggests that 
 one reason for adhering to binary-relevance measures
 is that an existing tool lacked the capability to handle graded relevance.
 On the other hand, this may not be the only reason:
 in the MathIR overview paper, it is reported that the organisers 
 chose Precision (a set retrieval measure, not a ranked retrieval measure)
 because it is ``\textit{simple to understand}''~\cite{zanibbi16}.
 Thus, 
 some researchers indeed \emph{choose} to focus on evaluation 
 with binary-relevance measures, even in the NTCIR community where we have 
 graded relevance data by default and a tool for 
 computing graded relevance measures is known\footnote{
 \texttt{NTCIREVAL}
  has been available on the NTCIR website since 2010;
 its predecessor \texttt{ir4qa\_eval} was released in 2008~\cite{sakai08ir4qa}.
 Note also that 
 TREC 2010 released
 \url{https://trec.nist.gov/data/web/10/gdeval.pl}
 for computing \ac{nDCG} and \ac{ERR}.
 }.

\section{Graded Relevance Assessments, Graded Relevance Measures}\label{s:graded}

This section provides an overview of NTCIR ranked retrieval tasks that 
employed  graded-relevance evaluation measures 
to fully enjoy the benefit of having graded relevance assessments.

\subsection{Web (NTCIR-3 through -5)}\label{ss:web}

The NTCIR-3 Web Retrieval task~\cite{eguchi03} 
was the first NTCIR task to use a graded relevance evaluation measure
namely, \ac{DCG}\footnote{
This was the DCG as originally defined by \citet{jarvelin00}
with the logarithm base $b=2$,
which means that gain discounting is not applied to documents at ranks 1 and~2.
That is, a SERP that has a relevant document at rank~1 
and one that has the same document at rank~2 are considered equally effective 
according to this measure. See also Section~\ref{ss:ir4qa}.
}.
Four-point relevance levels 
were used: 
highly relevant, fairly relevant, partially relevant, and nonrelevant.
In addition, assessors chose a very small number 
of ``best'' documents from the pools.
To compute DCG,
two different gain value settings were used:
\begin{description}
\item[Rigid] 3 for highly relevant, 2 for fairly relevant, 0 otherwise;
\item[Relaxed] 3 for highly relevant, 2 for fairly relevant,  1 for partially relevant, 0 otherwise.
\end{description}
The organisers of the Web Retrieval task 
also defined a graded-relevance evaluation measure
called \ac{WRR} (first mentioned in Section~\ref{sss:Qvariants}),
designed for navigational searches (which were called Target Retrieval in the Web task).
\ac{WRR} is an extension of the \ac{RR} measure,
and
 assumes, for example,
that having a marginally relevant document at rank~1 
is more important than having a highly relevant one at rank~2.
However, what was actually used in the task 
was the binary-relevance \ac{RR},
with two different thresholds for 
mapping the 4-point relevance levels into binary.
Therefore, this measure will be denoted ``(W)RR''
hereafter whenever graded relevance is not utilised.
Other binary-relevance measures including AP and R-precision 
were also used in this task.
For a comparison of evaluation measures designed 
for navigational intents, including \ac{RR}, \ac{WRR}, and the aforementioned 
P$+$, we refer the reader to \citet{sakai07tod}.

The NTCIR-14 WEB Informatinal Retrieval Task~\cite{eguchi04}
was similar to its predecessor, 
with 4-point relevance levels;
the evaluation measures were DCG, (W)RR, Precision etc.
On the other hand,
the NTCIR-14 WEB Navigational Retrieval Task~\cite{oyama04},
used 3-point relevance levels:
A (relevant), 
B (partially relevant),
and
D (nonrelevant);
the evaluation measures were DCG and (W)RR, and
two gain values settings for DCG were used:
$(A,B,D)=(3,0,0)$ and $(A,B,D)=(3,2,0)$.
Also at NTCIR-4, an evaluation measure for web search called 
the User's Character Score (UCS) was proposed~\cite{ohtsuka04},
which basicaly assumes that having relevant documents at consecutive ranks 
is better than having them alternately with nonrelevant ones.
However, this is a binary-relevance measure:
the proposers argue that
not requiring graded relevance assessments is an advantage.

The NTCIR-15 WEB task ran 
the Navigational Retrieval subtask,
which is basically the same as its predecessor,
with 3-point relevance levels and DCG and (W)RR.
For computing DCG, three gain value settings 
were used: 
$(A,B,D)=(3,0,0)$, $(A,B,D)=(3,2,0)$,
and $(A,B,D)=(3,3,0)$.
Note that the first and the third settings reduce DCG to binary-relevance measures.

\subsection{CLIR (NTCIR-6)}\label{ss:clir6}

At the NTCIR-6 CLIR task,
4-point relevance levels (S,A,B,C) were used 
and rigid and relaxed AP scores were computed using {\tt trec\_eval} as before.
In addition, the organisers 
computed ``\textit{as a trial}''~\cite{kishida07}
 the following graded-relevance measures using their own script:
\ac{nDCG} (as defined originally by \citet{jarvelin02}), Q-measure, and Kishida's \emph{generalised AP}~\cite{kishida05genap}.
See \citet{sakai07evia} for a comparison of these three graded-relevance measures.
The CLIR organisers developed a programme to compute these graded-relevance
measures, with the gain value setting: $(S,A,B,C)=(3,2,1,0)$.

\subsection{ACLIA IR4QA (NTCIR-7 and -8)}\label{ss:ir4qa}

At the NTCIR-7 \ac{IR4QA} task~\cite{sakai08ir4qa},
a predecessor of \texttt{NTCIREVAL} called \texttt{ir4qa\_eval} was released (See Section~\ref{ss:Q}).
This tool was used to compute the Q-measure,
the ``Microsoft version'' of \ac{nDCG}~\cite{burges05}, 
as well as the binary-relevance AP.
Microsoft nDCG (called \texttt{MSnDCG} in \texttt{NTCIREVAL}) fixes a problem with the original nDCG (See also Section~\ref{ss:web}):
in the original nDCG, if the logarithm base is set to (say) $b=10$,
then discounting is not applied from ranks 1 to 10,
and therefore nDCG at cutoff $l=10$ 
is reduced to nCG (i.e., NSR; See Section~\ref{sss:Q}) at $l=10$.
This is 
 a set retrieval measure
rather than a ranked retrieval measure;
the ranks of the relevant documents within top 10 do not matter.
Microsoft nDCG avoids this problem\footnote{
\citet{jarvelin08} describe another version of nDCG to fix the problem with the original nDCG.
}
 by 
using $1/\log(1+r)$ as the discount factor for \emph{every} rank $r$,
but thereby loses the patience parameter $b$~\cite{sakai14promise}\footnote{
D$\sharp$-nDCG implemented in \texttt{NTCIREVAL} also 
builds on the Microsoft version of nDCG, not the original nDCG.
}.
The relevance levels used were: 
L2 (relevant),
L1 (partially relevant),
and
L0 (nonrelevant);
A \emph{linear} gain value setting was used: $(L2, L1, L0)=(2,1,0)$.
The NTCIR-8 IR4QA task~\cite{sakai10ir4qa} used the same evaluation 
methodology as above.

\subsection{GeoTime (NTCIR-8 and -9)}

The NTCIR-8 GeoTime task~\cite{gey10}, which dealt with adhoc IR given 
``when and where''-type topics,
constructed test collections with 
the following graded relevance levels:
\begin{description}
\item[Fully relevant] The document answers both the ``when'' and ``where'' aspects of the topic;
\item[Partially relevant -- where] The document only answers the ``where'' aspect of the topic;
\item[Partially relevant -- when] The document only answers the ``when'' aspect of the topic.
\end{description}
The evaluation tools from the IR4QA task were used to compute
(Microsoft) nDCG, Q, and AP,
with a gain value of 2 for each fully relevant document
and a gain value of 1 for each partially relevant one (regardless of ``when'' or ``where'') for the two
graded-relevance measures\footnote{
While the GeoTime overview paper suggests
that the above relevance levels were mapped to binary relevance,
this was in fact not the case: it was the present author who conducted the official evaluation 
for both NTCIR-8 IR4QA and GeoTime; they were done in exactly the same way,
by utilising the 3-point relevane levels.
}.
The NTCIR-9 GeoTime task~\cite{gey11} used the same evaluation 
methodology as above.

\subsection{CQA (NTCIR-8)}

The NTCIR-8 \ac{CQA} task~\cite{sakai10cqa} was an answer ranking task:
given a question from Yahoo! Chiebukuro (Japanese Yahoo! Answers)
and the answers posted in response to that question,
rank the answers by answer quality.
While the Best Answers (BAs) selected by the actual questioners
were already available in the Chiebukuro data,
additional \emph{graded} relevance assessments were obtained offline
to find \emph{Good Answers} (GAs),
by letting four assessors independently judge each posted answer.
Each assessor labeled an answer
as either
A (high-quality),
B (medium-quality),
or 
C (low-quality),
and hence 15 different 
label \emph{patterns} were obtained: 
$AAAA, AAAB, \ldots, BCCC, CCCC$. 
In the official evaluation at NTCIR-8,
these patterns were mapped to 4-point relevance levels:
for example, $AAAA$ and $AAAB$ were mapped to $L3$-relevant,
and $ACCC, BCCC$ and $CCCC$ were mapped to $L0$.
In a separate study~\cite{sakai11wsdm},
the same data were mapped to 9-point relevance levels,
by giving 2 points to an A and 1 point to a B
and summing up the scores for each pattern.
That is, the score for $AAAA$ would be 8
and therefore the relevance level assigned is $L8$;
similarly, $AAAB$ is mapped to $L7$; both
$AABB$ and $AAAC$ are mapped to $L6$, and so on.
While this is similar to the approach
taken in the Chinese-English IR tasks of NTCIR-2~\cite{chen01},
recall that the NTCIR-2 task did not utilise any graded relevance 
measures (Section~\ref{ss:earlyIR}).

Using the graded Good Answers data,
three graded-relevance measures were computed:
normalised gain at $l=1$ (nG@1)\footnote{
nG@1 is often referred to as nDCG@1; however,
note that neither discounting nor cumulation is applied at rank 1.
},
nDCG, 
and Q.
In addition, Hit at $l=1$ (defined as ${\it Hit}@1=I(1)$ using the relevance flag from Section~\ref{sss:Q})
was computed for both Best Answers 
and Good Answers data:
this is a binary-relevance measure
which only cares whether the top ranked item is relevant or not.

The CQA organisers also experimented with 
an attempt at constructing binary-relevance assessments
based on the Good Answers data.
For each assessor,
answers that were most highly rated by that assessor among all the posted answers
were identified as his/her ``favourite'' answers; 
note that they may not necessarily be rated $A$.
Then the union of the favourite answers from all four assessors
were treated as relevant. Furthermore,
the best answer selected by the questioner was also 
added to the binary-relevance set,
for the best answer was in fact the questioner's favourite answer.

\subsection{INTENT/IMine (NTCIR-9 through 12)}\label{ss:intent}

The NTCIR-9 INTENT task overview paper~\cite{song11} was the first 
NTCIR overview to mention the use of the \texttt{NTCIREVAL} tool,
which can compute various graded-relevance measures
for adhoc and diversified IR, including the Q-measure, nDCG, and
D$\sharp$-measures.
D$\sharp$-nDCG and its components I-rec and D-nDCG
were used as the official evaluation measures.
The Document Retrieval (DR) subtask of the INTENT
task had intentwise graded relevance assessments on 
a 5-point scale: $L0$ through $L4$.
This was obtained by hiring two assessors per topic,
who independently judged each document 
as either highly relevant or relevant.

While the Subtopic Mining (SM) subtask of the INTENT task
also used D$\sharp$-nDCG 
to evaluate ranked lists of subtopic strings,
no graded relevance assessments were involved in the SM subtask
since each subtopic string either belongs
to an intent (i.e., a cluster of subtopic strings) or not.
Hence, the SM subtask may be considered to be outside the scope
of the present survey.
However, there is an interesting aspect to the evaluation of the SM subtask
when D$\sharp$-nDCG is used, from a graded relevance point of view.
Recall the definition of the global gain (Eq.~\ref{eq:GG}):
when the intentwise relevance assessments are binary as is the case here,
the global gain is reduced to $\sum_{i} {\it Pr}(i|q)$, 
i.e., the sum of the intent probabilities.
Furthermore, 
since relevant subtopic string belongs to exactly one intent in the SM subtask,
the global gain of a subtopic string that belongs to intent~$i$
is given exactly by ${\it Pr}(i|q)$, which is estimated based on the number of votes from assesors.
That is, a D-nDCG score for the SM subtask is exactly an nDCG score
where each gain value is given by the probability of the intent
to which the subtopic string belongs.

In both SM and DR subtasks,
the trade-offs between D-nDCG (i.e., overall relevance)
and I-rec (pure diversity) were visualised in the overview paper.

The NTCIR-10 INTENT task was basically the same 
as its predecessor, with 5-point intentwise relevance levels for the DR subtask
and D$\sharp$-nDCG as the primary evaluation measure.
However,
as the intents came with informational/navigational tags,
DIN-nDCG and P$+$Q were also used to leverage this information (See Section~\ref{sss:Dvariants}).

The NTCIR-11 IMine task\cite{liu14} was similar to the INTENT tasks,
except that its SM subtask 
required participating systems to return a two-level hierarchy of subtopic strings.
As was described in Section~\ref{sss:Dvariants},
the SM subtask was evaluated using the H-measure,
which combines (a)~the accuracy of the hierarchy (i.e.,
whether the second-level subtopics are correctly assigned to the first-level ones),
(b)~the D$\sharp$-nDCG score based on the ranking of the first-level subtopics,
and (c)~the D$\sharp$-nDCG score based on the ranking of the second-level subtopics.
However, recall the above remark on the INTENT SM subtask:
intentwise graded relevance does not come into play in this subtask.
On the other hand, the IMine DR subtask was evaluated in a way similar to 
the INTENT DR tasks, with D$\sharp$-nDCG computed 
based on 4-point relevance levels:
highly relevant, relevant, nonrelevant, and spam.
The gain value setting used was: $(2,1,0,0)$\footnote{
Kindly confirmed by task organisers Yiqun Liu and Cheng Luo
in a 
private email communication  (March 2019).
}.

The IMine task also introduced the TaskMine subtask,
which requires systems to rank strings that represent subtasks
of a given task (e.g., ``\textit{take diet pills}'' in response to ``\textit{lose weight}.'')
This subtask involved graded relevance assessments.
Each subtask string was judged independently by two assessors
from the viewpoint of whether the subtask is effective for achieving the input task.
A 4-point per-assessor relevance scale was used\footnote{
While the overview (Section 4.3) says that a 3-point scale was used, this was in fact not the case:
kindly confirmed by task organiser Takehiro Yamamoto in a private email communication (March 2019).
}, with weights $(3,2,1,0)$,
and final relevance levels were given as the average of the two scores,
which means that a 6-point relevance scheme was adopted.
The averages were used verbatim as gain values:
$(3.0, 2.5, 2.0, 1.5, 1.0, 0)$.
The evaluation measure used was nDCG,
but duplicates (i.e. multiple strings representing the same subtask)
were not rewarded,
just as in the original Q-measure (See Section~\ref{sss:Q})
and the NTCIR Web tasks~\cite{eguchi03}.

As was already discussed in Section~\ref{sss:Dvariants},
the Query Understanding (QU) subtask of the NTCIR-12 IMine-2 Task~\cite{yamamoto16},
a successor of the previous SM subtasks of INTENT/IMine,
required systems 
to return a ranked list of $($subtopic, vertical$)$ pairs (e.g., $($``\textit{iPhone 6 photo}'', Image$)$, 
$($``\textit{iPhone 6 review}'', Web))
for a given query.
The official evaluation measure, called the QU-score (Eq.~\ref{eq:QU}),
is a linear combination of D$\sharp$-nDCG (computed as in the INTENT SM subtasks)
and the V-score (Eq.~\ref{eq:Vscore})
which measures the appropriateness of the named vertical for each subtopic string.
Despite the binary-relevance nature of the subtopic mining aspect of the QU subtask,
it deserves to be discussed in the present survey because
the V-score part relies on graded relevance assessments.
To be more sepcific, the V-score
relies on the probabilities $\{{\it Pr}(v|i)\}$,
for intents $\{i\}$ and verticals $\{v\}$,
which are derived from
3-point scale relevance assessments:
2 (highly relevant), 
1 (relevant), 
and 
0 (nonrelevant).
Hence the QU-score may be regarded as a graded relevance measure.

The Vertical Incorporating (VI) subtask of the NTCIR-12 IMine-2 Task~\cite{yamamoto16}
also used a version of D$\sharp$-nDCG to allow systems 
to embed verticals (e.g., Vertical-News, Vertical-Image) within a ranked list of document IDs for diversified search.
More specifically, the organisers 
replaced the intentwise gain value $g_{i}(r)$ at rank $r$ in the global gain formula
(Eq.~\ref{eq:GG}) with 
${\it Pr}(v(r)|i) g_{i}(r)$,
where $v(r)$ is the \emph{vertical type} 
(``Web,''  
Vertical-News, Vertical-Image, etc.)
of the document at rank $r$, and the vertical probability given an intent
is obtained from 3-point relevance assessments as described above.
As for the intentwise gain value $g_{i}(r)$,
it was 
also on a 3-point scale for the Web documents:
2 for highly relevant, 1 for relevant, and 0 for nonrelevant documents.
Moreover, if the document at $r$ was a vertical,
the gain value was set to 2.
That is, the verticals were treated as highly relevant.
In addition, the VI subtask 
collected \emph{topicwise} relevance assessments
on a 4-point scale: 
highly relevant, relevant, nonrelevant, and spam.
The gain values used were:
$(2,1,0,0)$\footnote{
Kindly confirmed by task organisers Yiqun Liu and Cheng Luo in a 
private email communication (March 2019).
}.
As the subtask had a set of \emph{very clear}, single-intent topics among their full topic set,
Microsoft nDCG (rather than D$\sharp$-nDCG) was used for these particular topics.

\subsection{RecipeSearch (NTCIR-11)}

The NTCIR-11 RecipeSearch Task~\cite{yasukawa14}
had two subtask:
\emph{Adhoc Recipe Search} and \emph{Recipe Pairing}.
Hereafter, we shall only discuss the former,
as the latter only had binary relevance assessments.
While the official evaluation results of Adhoc Recipe Search
was based on binary relevance,
the organisers also explored evaluation based on graded relevance:
they obtained graded relevance assessments on a 3-point scale (L2, L1, L0)
for a subset (111 topics) of the full test topic set (500 topics)\footnote{
While the overview paper says that a 4-point scale was used,
this was in fact not the case: kindly confirmed by 
task organiser Michiko Yasukawa (March 2019) in a private
email communication.
}.
Microsoft nDCG was used to leverage the above data
with a linear gain value setting,
along with the binary \ac{AP} and \ac{RR}.

\subsection{Temporalia (NTCIR-11 and -12)}

The Temporal Information Retrieval (TIR) subtask
of the NTCIR-11 Temporalia Task collected 
relevance assessments on a 3-point scale:
highly relevant, relevant, and nonrelevant.
Each TIR topic
contained a \emph{past question},
\emph{recency question},
\emph{future question}, and an
\emph{atemporal question},
in addition to the description and search date fields;
participating systems were required to produce a \ac{SERP}
for each of the above four questions.
This adhoc IR task used Precision
and Microsoft nDCG as the official measures,
and the Q-measure for reference. 

While the Temporally Diversified Retrieval (TDR) subtask
of the NTCIR-12 Temporalia-2 Task was similar to the above 
TIR subtask,
it required systems to return a fifth SERP,
which covers all of the above four temporal classes.
That is, this fifth SERP is a diversified SERP,
where the temporal classes can be regarded as different search intents 
for the same topic.
The relevance assessment process followed the practice of the NTCIR-11 TIR task
(with 3-point relevance levels), and the SERPs to the four questions 
were evaluated using nDCG.
As for the diversified SERPs, they were 
evaluated using $\alpha$-nDCG~\cite{clarke08} and D$\sharp$-nDCG.

A linear gain value setting was used in both of the above subtasks\footnote{
Kindly confirmed by task organiser Hideo Joho in a private email communication (March 2019).
}.

\subsection{STC (NTCIR-12 through -14)}\label{ss:STC}

The NTCIR-12 \ac{STC} task~\cite{shang16} was basically 
a response retrieval task given a tweet (or a Chinese Weibo post; both will be referred to generically as ``tweet'' hereafter).
For both Chinese and Japanese subtasks,
the response tweets
were first labelled on a binary scale,
for each of the following criteria:
\begin{description}
\item[Coherence] Does the response logically make sense as a response to the input tweet?
\item[Topical relevance] Does the topic of the response match that of the input tweet?
\item[Context independence] Is the response appropriate for the input tweet regardless of the outisde context?
\item[Non-repetitiveness] Does the response contain something new, not just a simple repetition of the input tweet?
\end{description}
The final graded relevance levels were determined using the following mapping scheme:
\begin{tabbing}
{\bf if} \=Coherent AND Topically Relevant\\
          \>{\bf if} \= Context-independent AND Non-repetitive\\
          \>			 \> RelevanceLevel = $L2$\\
          \>{\bf else}\\
          \>			\> RelevanceLevel = $L1$\\
{\bf else}\\
		   \> RelevanceLevel = $L0$.\\
\end{tabbing}
Following the \emph{quadratic} gain value setting often used for web search evaluation~\cite{burges05}
and for computing \ac{ERR}~\cite{chapelle09},
the Chinese subtask organisers mapped 
the L2, L1, and L0 relevance levels to the following gain values:
$2^{2}-1=3, 2^{1}-1=1, 2^{0}-1=0$;
according to the present survey of NTCIR retrieval tasks,
this is the only case where a quadratic gain value setting was used 
instead of the linear one.
The evaluation measures used for this subtask were nG@1,
P$+$, and normalised \ac{ERR} (nERR).
As for the Japanese subtask which used Japanese Twitter data,
the same mapping scheme was applied, 
but the scores ($(L2,L1,L0)=(2,1,0)$) from
10 assessors were averaged 
to determine the final gain values;
a binary-relevance, set-retrieval accuracy measure 
was used instead of P$+$, along with nG@1 and nERR.

The NTCIR-13 \ac{STC} task~\cite{shang17} was similar to its predecessor,
although systems were allowed to \emph{generate} responses intead of retrieving existing tweets.
The Chinese subtask used the following new criteria and the mapping scheme to obtain per-assessor graded relevance scores:
\begin{description}
\item[Fluent] The response is acceptable as a natural language text;
\item[Coherent] The response is logically connected and topically relevant to the input post;
\item[Self-sufficient] The assessor can judge that the response is appropriate by reading just the post-response pair;
\item[Substantial]  The response provides new information in the eye of the author of the input post.
\end{description}
\begin{tabbing}
{\bf if} \=Fluent AND Coherent\\
          \>{\bf if} \= Self-sufficient AND Substantial\\
          \>			 \> AssessorScore = $2$\\
          \>{\bf else}\\
          \>			\> AssessorSore = $1$\\
{\bf else}\\
		   \> AssessorScore = $0$.\\
\end{tabbing}
Finally, 7-point relevance levels L0 through L6 were obtained by summing up the
assessor scores, and a linear gain value setting was used to compute 
nG@1, P$+$, and nERR.
In addition, an alternative approach to consolidating the assessor scores
was explored, by considering the fact that
some receive unanimous ratings while others do not 
even if they are the same in terms of the sum of assessor scores.
More specifically, 
the raw gain value ${\it gv}$ (i.e., sum of the assessor scores)
was ``upgraded'' based on \emph{unanimity} as follows~\cite{sakai17evia-unan}:
\begin{equation}
{\it ugv} = {\it gv} + p N (D_{\it max}-D) \ ,
\end{equation}
where $p$ is an upgrade strength parameter (set to $p=0.2$),
$N$ is the number of assessors ($N=3$ for the subtask),
$D_{\it max}$ is the highest possible assessor score ($D_{\it max}=2$ in this case),
and $D$ is the difference between the highest and the lowest assessor scores 
for the item in question.
For example,
while a response labelled  $(2,1,1)$ and another labelled $(2,2,0)$
would receive the same raw gain value of 4,
the unanimity-aware gain values would be 4.6 and 4.0, respectively.

The NTCIR-13 \ac{STC} Japanese subtask
used Yahoo! News Comments data (user interactions on an online news article page)
 instead of Japanese Twitter data.
Accordingly, the following new criteria for obtaining per-assessor scores
were used:
\begin{description}
\item[Fluent] The response is fluent and understandable from a grammatical point of view (possible scores: 1,0);
\item[Coherent] The response maintains coherence with the news topic and the input comment (possible scores: 1,0);
\item[Context-dependent] The response depends on and is related to the input comment (possible scores: 2,1,0)\footnote{
This criterion is different from Context-\emph{independence} used in the NTCIR-12 STC task.
};
\item[Informative] The response is informative and influences the author of the comment (possible scores: 2,1,0).
\end{description}
Note that the Context-dependence and Informativeness criteria are not binary.
The Japanese subtask used the following two different schemes to map the scores to per-assessor scores:
\begin{tabbing}
SCHEME1:\\
{\bf if} \=Fluent AND Coherent\\
          \>{\bf if} \= Context-dependent == 2 AND Informative == 2\\
          \>			 \> AssessorScore = $2$\\
          \>{\bf else}\\
          \>			\> AssessorSore = $1$\\
{\bf else}\\
		   \> AssessorScore = $0$.\\
	\\
SCHEME2:\\
{\bf if} \=Fluent AND Coherent\\
          \>{\bf if} \= Context-dependent == 2 AND Informative == 2\\
                    \>			 \> AssessorScore = $2$\\
          \>{\bf else if} Context-dependent == 0 OR Informative ==0\\
                    \>			 \> AssessorScore = $0$\\
          \>{\bf else}\\
          \>			\> AssessorSore = $1$\\
{\bf else}\\
		   \> AssessorScore = $0$.\\
\end{tabbing}
Five assessors independently judged each response 
and the per-assessor scores were averaged to compute the evaluation measures:
nG@1, nERR, and the binary accuracy.

Although the Chinese Emotional Conversation Generatin (CECG) subtask
of the NTCIR-14 \ac{STC} subtask~\cite{zhang19} is not exactly a ranked retrieval task,
we discuss it here as it is a successor of the previous Chinese STC subtasks
and utilises graded relevance measures.
Given an input tweet \emph{and an emotional category} such as Happiness and Sadness,
participating systems for this subtask were required to return \emph{one} generated response.
In addition to the aforrementioned Fluency and Coherence criteria,
assessors were asked to judge whether the returned response 
is consistent with the emotional category specified in the input.
The following mapping scheme was used to determine per-assessor relevance levels:
\begin{tabbing}
{\bf if} \=Fluent AND Coherent\\
          \>{\bf if} \= Emotion-Consistent\\
          \>			 \> RelevanceLevel = $L2$\\
          \>{\bf else}\\
          \>			\> RelevanceLevel = $L1$\\
{\bf else}\\
		   \> RelevanceLevel = $L0$.\\
\end{tabbing}
The above relevance levels from
three crowd workers were consolidated on a majority-vote basis,
but if all three labels differed from one other (i.e., L2 vs. L1 vs. L0),
the final relevance level was set to L0.
As for the evaluation measures,
the relevance scores $(L2,L1,L0)=(2,1,0)$ of the returned responses
 were
simply summed or averaged across the test topics.

\subsection{WWW (NTCIR-13 and -14) and CENTRE (NTCIR-14)}

The NTCIR-13 We Want Web (WWW) Task~\cite{luo17} was an adhoc web search task.
For the Chinese subtask,
three assessors independently judged each pooled web page
on a 4-point scale: highly relevant (3 points), relevant (2 points), marginally relevant (1 point), and nonrelevant (0 points);
the scores were then summed up to form the final 10-point relevance levels, L0 through L9.
For the English subtask,
two assessors indepenently judged each pooled web page 
on a different 4-point scale: 
highly relevant (2 points), relevant (1 point), nonrelevant (0 points), 
and error (the web page from \texttt{clueweb12-B13}\footnote{
\texttt{http://lemurproject.org/clueweb12/}
} 
could not be displayed properly on the judgement interface; also 0 points);
the scores were then summed up to form the final 5-point relevance levels, L0 through L4.
In both subtasks, linear gain value settings were used to compute (Microsoft) nDCG, Q (the cutoff-based version given by
Eq.~\ref{eq:Qcutoff}), and nERR.

The NTCIR-14 WWW Task~\cite{mao19} was similar to its predecessor.
The Chinese subtask
used the following judgment criteria:
highly relevant (3 points), relevant (2 points), marginally relevant (1 point),
nonrelevant (0 points), garbled (similar to ``error'' mentioned above; also 0 points).
Although three assessors judged each topic,
the final relevance levels were obtained on a majority-vote basis rather than taking the sum;
hence 4-point scale relevance levels L0 through L3 were used this time.
As for the English subtask, 5-point relevance levels were obtained by following the 
methodology of the NTCIR-13 English subtask.
Both subtasks adhered to  Microsoft nDCG, (cutoff-based) Q, and nERR
with linear gain value settings.

The NTCIR-14 \ac{CENTRE} task~\cite{sakai19centre}
encouraged participants 
to replicate a pair of runs from the NTCIR-13 WWW English subtask
and to reproduce a pair of runs from the TREC 2013 Web Track adhoc task~\cite{collins-thompson14}.
Additional relevance assessments were conducted on top of the official NTCIR-13 WWW English 
test collection, by following the relevance assessment methodology of the WWW subtask.
As for the evaluation of the TREC runs with the TREC 2013 Web Track adhoc test collection,
the original 6-point scale relevance levels
Navigational, Key, Highly relevant, Relevant, Nonrelevant, Junk
were mapped to $L4, L3, L2, L1, L0, L0$ respectively.
All runs involved in the CENTRE task were evaluated using  Microsoft nDCG, (cutoff-based) Q, and nERR,
with linear gain value settings.

\subsection{AKG (NTCIR-13)}

The NTCIR-13 \ac{AKG} task~\cite{blanco17} had two subtasks:
\emph{Action Mining} (AM) and \emph{Actionable Knowledge Graph Generation} (AKGG).
Both of them involved graded relevance assessments and graded relevance measures.
The AM subtask required systems to rank \emph{actions} for 
a given \emph{entity type} and an \emph{entitiy instance}:
for example, 
given ``\textit{Product}'' and ``\textit{Final Fantasy VIII},'' 
the ranked actions could contain ``\textit{play on Android},'' ``\textit{buy new weapons}'' etc.
Two sets of relevance assessments were collected by means of crowdsourcing:
the first set judged the verb parts of the actions (``\textit{play},'', ``\textit{buy}'' etc.)
whereas the second set judged the entire actions (verb plus modifier as exemplified above).
Both sets of judgements were done based on 4-point relevance levels: L0 through L3.
The AKGG subtask required participants to rank entity \emph{properties}:
for example, given a quadruple (Query, Entity, Entity Types, Action)=(``\textit{request funding},''
``\textit{funding},'' 
``\textit{thing, action},''
``\textit{request funding}''),
systems might return ``\textit{Agent},'' ``\textit{ServiceType},'' ``\textit{Result}'' etc.
Relevance assessments were conducted by crowd workers on a 5-point scale:
L4 (Perfect), L3 (Excellent), L2 (Good), L1 (Fair), and L0 (Bad).
Both subtasks used nDCG and nERR for the evaluation.
Again, linear gain value settings were used\footnote{
Kindly confirmed by task organiser Hideo Joho in a private email communication (March 2019).
}.

\subsection{OpenLiveQ (NTCIR-13 and -14)}

The NTCIR-13 OpenLiveQ task~\cite{kato17}
required participants to rank Yahoo! Chiebukuro questions
for a given query, and the offline evaluation part of this task 
involved ranked list evaluation with graded relevance.
Five crowd workers independently judged a list of questions for query $q$
under the following instructions:
``Suppose you input $q$ and received a set of questions as shown below.
Please select all the questions that you would want to click.''
Thus, while the judgement is binary for each assessor,
6-point relevance levels (L0 through L5) were obtained based on the number of votes.
(Microsoft) nDCG, Q, and ERR were computed using a linear gain value setting.

The NTCIR-14 OpenLiveQ-2 task~\cite{kato19}
is similar to its predecessor,
but this time the evaluation involved \emph{unjudged} documents,
as the relevance assessments from NTCIR-13 were reused 
but the target questions to be ranked were not identical to the NTCIR-13 version.
The organisers therefore used \emph{condenlised-list}~\cite{sakai07sigir,sakai08irj} versions 
of Q, (Microsoft) nDCG,  and ERR: that is, the measures
are computed after removing all unjudged questions from the ranked lists.
Also, for OpenLiveQ-2, the organisers
switched their primary measure from nDCG to Q,
as Q substantially outperformed nDCG (at $l=5,10,20$)
in terms of correlation with online (i.e., click-based) 
evaluation in their experiments~\cite{kato18}.

\section{Summary}\label{s:summary}

\begin{table}[t]
\caption{NTCIR Ranked Retrieval Tasks with graded relevance assessments and binary relevance measures.
Note that the relevance levels for the Patent Retrivals tasks of NTCIR-4 to -6 exclude the ``nonrelevant'' level:
the actual labels are shown here because they are not simply different degrees of relevance (See Section~\ref{ss:patent}).
}
\label{t:binary}      

\begin{tabular}{llll}

\hline
Task or Subtask & NTCIR &Relevance levels &Main evaluation measures\\
       & round (year) &                       &discussed in overview\\
\svhline
Japanese and 		&1 (1999)				&3		& AP, R-precision, Precision,\\
\hspace*{2mm}JEIR	&						&			&RP curves\\
JEIR					&2	(2001)				&4		&AP, R-precision, Precision,\\
						&								&			&Interpolated Precision,\\
						&								&			&RP curves\\
Chinese and CEIR	&2					&4 per assessor	&RP curves\\
CLIR					&3-5							&4		&AP, RP curves\\
						&(2002-2005)			&			&\\
Patent Retrieval		&3 (2002)			&4		&RP curves\\
Patent Retrieval		&4 (2004)			&A,B 	&AP, RP curves\\
Patent Retrieval		&5 (2005)			&A,B 	&CRS (for passage retrieval), AP\\
Patent Retrieval		&6 (2007)			&A,B / H,A,B (Japanese)	&AP\\
								&						&A,B (English)					&AP\\
Spoken Document/	&9-11 				&3	&AP and passage-level variants\\
\hspace*{2mm}Content Retrieval								&(2011-2014)	&				&\\
SQ-SCR (SGS)	&12 (2016) 			&3	&AP\\									
Math Retrieval			&10 (2013)		&5 mapped to 3		&AP, Precision\\
Math Retrieval			&11 (2014)		&3					&AP, Precision, Bpref\\
MathIR						&12 (2016)		&3					&Precision\\
\hline
\end{tabular}
\end{table}

\begin{table}[t]
\caption{NTCIR Ranked Retrieval Tasks with graded relevance assessments and graded relevance measures.
Binary-relevance measures are shown in parentheses.}
\label{t:graded}

\begin{tabular}{llll}

\hline
Task or Subtask & NTCIR &Relevance levels & Main evaluation measures\\
       & round (year) &                       &discussed in overview\\
\svhline
Web Retrieval					&3 (2003)					&4 $+$ best documents & DCG ((W)RR, AP, RP curves)\\
WEB Informational			&4 (2004)			&4	&DCG ((W)RR, Precision, RP curves)\\
WEB Navigational				& 				&3	&DCG, ((W)RR, UCS)\\
WEB Navigational				&5 (2005)		&3		&DCG, ((W)RR)\\
CLIR			&6	(2007)						&4				&nDCG, Q, generalised AP (AP)\\
IR4QA					&7-8						&3			&nDCG, Q (AP)\\
				&(2008-2010)					&				&\\
GeoTime		&8-9								&3$\ast$		&nDCG, Q (AP)\\
				&(2010-2011)					&					&\\
CQA			&8	(2010)						&4(9) $+$ best answers 		&GA-\{nG@1, nDCG, Q\}, \\
				&										&			&(GA-Hit@1, BA-Hit@1) etc.\\
INTENT DR	&9 (2011)						&5			&D$\sharp$-nDCG\\
INTENT DR	&10 (2013)					&5			&D$\sharp$-nDCG, DIN-nDCG, P$+$Q\\
IMine  DR	    &11 (2014)					&4 incl. Spam 			&D$\sharp$-nDCG\\
IMine TaskMine&11 							&6 			&nDCG\\
IMine QU		&12 (2016)					&3 (vertical)&QU-score\\
IMine VI	&12								&3 (vertical) &D$\sharp$-nDCG, nDCG\\
  				& 								&3 (intentwise)&\\
  				&								&3 $+$ Spam (topicwise)&\\
RecipeSearch	&11 (2014)					&3(2)		&nDCG (AP, RR)\\
Temporalia TIR	&11						&3					&nDCG, Q, (Precision)\\
Temporalia TDR	&12 (2016)				&3 					&nDCG, $\alpha$-nDCG, D$\sharp$-nDCG\\
STC Chinese &12	 (2016)					&3		&nG@1, P$+$, nERR\\
STC Chinese &13 (2017)					&7(10)		&nG@1, P$+$, nERR\\
STC Japanese &12-13						&3 per assessor			&nG@1, nERR (Accuracy)\\	
STC CECG	&14 (2019)					&3			&sum/average of relevance scores\\						
WWW English		&13-14	&5				&nDCG, Q, nERR\\
					&(2017-2019)	&			&\\
WWW Chinese	&13 (2017)		&10	&nDCG, Q, nERR\\
WWW Chinese	&14				&4		&nDCG, Q, nERR\\
AKG 			&13 (2017)		&4 (AM) / 5 (AKGG)			&nDCG, nERR\\
OpenLiveQ	&13-14	&6					&nDCG, Q, ERR\\
				&(2017-2019)	&				&(with condensed lists at NTCIR-14)\\
CENTRE		&14 (2019)		&5			&nDCG, Q,nERR\\
\hline
\end{tabular}\\
$\ast$two types of partially relevant (\emph{when} and \emph{where}) counted as one level.
\end{table}

Table~\ref{t:binary} summarises the ranked retrieval tasks of NTCIR discussed in Section~\ref{s:binary},
i.e., those that used binary relevance evaluation measures even though they collected graded relevance 
assessments.
Similarly,
Table~\ref{t:graded} summarises the tasks discussed in Section~\ref{s:graded},
which fully utilised their graded relevance assessments by means of graded relevance measures.
It can be observed that:
\begin{itemize}
\item The majority of the past NTCIR ranked retrieval tasks, though not all of them,
utilised graded relevance measures;
\item Even a few relatively recent tasks, namely, SpokenQueryDoc and MathIR from NTCIR-12 held in 2016,
refrained from using graded relevance measures.
\end{itemize}

As was discussed in Section~\ref{ss:earlyIR},
researchers should be aware that
binary-relevance measures with different relevance thresholds
(e.g. Relaxed AP and Rigid AP)
cannot serve as substitutes for 
good graded-relevance measures.
\emph{If} 
the optimal ranked output for a task
is defined as one that sorts all relevant documents
 in decreasing order of relevance levels,
then by definition, graded relevance measures should be used to
evaluate and optimise the retrieval systems.

One additional remark regarding Tables~\ref{t:binary} and~\ref{t:graded}
is that
the NTCIR-5 CLIR overview paper~\cite{kishida07} was the last
to report on RP curves; 
the RP curves completely disappeared from the NTCIR overviews after that.
This may be because 
(a)~interpolated precisions at different recall points~\cite{sakai14promise} do not directly reflect user experience; and
(b)~graded-relevance measures have become more popular than before.

What lies beyond graded relevance? Here is my personal view
concerning offline evaluation (as opposed to online evaluation using click data etc.).
\ac{IR} and \ac{IA} tasks have diversified,
and relevance assessments require more subjective and diverse views
than before. We are no longer just talking about
whether a scientific article is relevant to the researcher's question (as in Cranfield);
we are also talking about whether a response of a chatbot
is ``relevant'' response to the user's utterance,
about whether a reply to a post on social media is ``relevant,'' and so on.
Graded relevance implies that there should be a \emph{single} label
for each item to be retrieved (e.g., ``this document is highly relevant''), 
but these new tasks may require a \emph{distribution} of labels
reflecting different users's points of view.
Hence, instead of collapsing this distribution to form a single label,
methods to preserve the distribution of labels in the test collection
may become useful.
The Dialogue Quality (DQ) and Nugget Detection (ND) subtasks of
 the NTCIR-14 STC task are the very first of NTCIR efforts in that direction;
 see also \citet{higashinaka17},
 \citet{maddalena17},
 and
 \citet{sakai18short}.

%
%

\input{acronyms-general}

\input{acronyms-sakai}

\bibliographystyle{spbasic}

\bibliography{ntcir-arxiv}

\end{document}

%% file: acronyms-general.tex
\acrodef{AKG}[AKG]{Actionable Knowledge Graph}
\acrodef{AP}[AP]{Average Precision}
\acrodef{CENTRE}[CENTRE]{CLEF NTCIR TREC Reproducibility}
\acrodef{CLEF}[CLEF]{Conference and Labs of the Evaluation Forum}
\acrodef{CQA}[CQA]{Community Question Answering}
\acrodef{DCG}[DCG]{Discounted Cumulative Gain}
\acrodef{ERR}[ERR]{Expected Reciprocal Rank}
\acrodef{FIRE}[FIRE]{Forum for Information Retrieval Evaluation}
\acrodef{IA}[IA]{Information Access}
\acrodef{IPC}[IPC]{International Patent Classification}
\acrodef{IR}[IR]{Information Retrieval}
\acrodef{IR4QA}[IR4QA]{Information Retrieval for Question Answering}
\acrodef{MAP}[MAP]{Mean Average Precision}
\acrodef{nCG}[nCG]{Normalized Cumulative Gain}
\acrodef{nDCG}[nDCG]{Normalized Discounted Cumulative Gain}
\acrodef{NTCIR}[NTCIR]{NACSIS Test Collection for Information Retrieval/NII Testbeds and Community for Information access Research}
\acrodef{OHSUMED}[OHSUMED]{Oregon Health Sciences University's MEDLINE Data Collection}
\acrodef{RR}[RR]{Reciprocal Rank}
\acrodef{QAC}[QAC]{Question Answering Challenge}
\acrodef{SERP}[SERP]{Search Engine Result Page}
\acrodef{STC}[STC]{Short Text Conversation}
\acrodef{TREC}[TREC]{Text REtrieval Conference}

%% file: acronyms-sakai.tex
\acrodef{NCP}[NCP]{Normalised Cumulative Precision}
\acrodef{NCU}[NCU]{Normalised Cumulative Utility}
\acrodef{PRP}[PRP]{Probability Ranking Principle}
\acrodef{WRR}[WRR]{Weighted Reciprocal Rank}

%% file: ntcir-arxiv.bbl
\begin{thebibliography}{93}
\providecommand{\natexlab}[1]{#1}
\providecommand{\url}[1]{{#1}}
\providecommand{\urlprefix}{URL }
\expandafter\ifx\csname urlstyle\endcsname\relax
  \providecommand{\doi}[1]{DOI~\discretionary{}{}{}#1}\else
  \providecommand{\doi}{DOI~\discretionary{}{}{}\begingroup
  \urlstyle{rm}\Url}\fi
\providecommand{\eprint}[2][]{\url{#2}}

\bibitem[{Agrawal et~al(2009)Agrawal, Gollapudi, Halverson, and
  Ieong}]{agrawal09}
Agrawal R, Gollapudi S, Halverson A, Ieong S (2009) Diversifying search
  results. In: Proceedings of ACM WSDM 2009, pp 5--9

\bibitem[{Aizawa et~al(2013)Aizawa, Kohlhase, and Ounis}]{aizawa13}
Aizawa A, Kohlhase M, Ounis I (2013) {NTCIR}-10 {Math} pilot task overview. In:
  Proceedings of NTCIR-10, pp 654--661

\bibitem[{Aizawa et~al(2014)Aizawa, Kohlhase, and Ounis}]{aizawa14}
Aizawa A, Kohlhase M, Ounis I (2014) {NTCIR}-11 {Math}-2 task overview. In:
  Proceedings of NTCIR-11, pp 88--98

\bibitem[{Akiba et~al(2011)Akiba, Nishizaki, Aikawa, Kawahara, and
  Matsui}]{akiba11}
Akiba T, Nishizaki H, Aikawa K, Kawahara T, Matsui T (2011) Overview of the
  {IR} for {Spoken} {Documents} task in {NTCIR}-9 workshop. In: Proceedings of
  NTCIR-9, pp 223--235

\bibitem[{Akiba et~al(2013)Akiba, Nishizaki, Aikawa, Hu, Ito, Kawahara,
  Nakagawa, Nanjo, and Yamashita}]{akiba13}
Akiba T, Nishizaki H, Aikawa K, Hu X, Ito Y, Kawahara T, Nakagawa S, Nanjo H,
  Yamashita Y (2013) Overview of the {NTCIR}-10 {SpokenDoc}-2 task. In:
  Proceedings of NTCIR-10, pp 573--587

\bibitem[{Akiba et~al(2014)Akiba, Nishizaki, Nanjo, and Jones}]{akiba14}
Akiba T, Nishizaki H, Nanjo H, Jones GJ (2014) Overview of the {NTCIR}-11
  {SpokenQuery\&Doc} task. In: Proceedings of NTCIR-11, pp 350--364

\bibitem[{Akiba et~al(2016)Akiba, Nishizaki, Nanjo, and Jones}]{akiba16}
Akiba T, Nishizaki H, Nanjo H, Jones GJ (2016) Overview of the {NTCIR}-12
  {SpokenQuery\&Doc}-2 task. In: Proceedings of NTCIR-12, pp 167--179

\bibitem[{Amig\'{o} et~al(2018)Amig\'{o}, Spina, and de~Albornoz}]{amigo18}
Amig\'{o} E, Spina D, de~Albornoz JC (2018) An axiomatic analysis of diversity
  evaluation metrics: Introducting the rank-biased utility metric. In:
  Proceedings of ACM SIGIR 2018, pp 625--634

\bibitem[{Blanco et~al(2017)Blanco, Joho, Jatowt, Yu, and Yamamoto}]{blanco17}
Blanco R, Joho H, Jatowt A, Yu H, Yamamoto S (2017) Overview of {NTCIR}-13
  {Actionable Knowledge Graph} ({AKG}) task. In: Proceedings of NTCIR-13, pp
  340--345

\bibitem[{Broder(2002)}]{broder02}
Broder A (2002) A taxonomy of web search. SIGIR Forum 36(2):3--10

\bibitem[{Buckley and Voorhees(2004)}]{buckley04}
Buckley C, Voorhees EM (2004) Retrieval evaluation with incomplete information.
  In: Proceedings of ACM SIGIR 2004, pp 25--32

\bibitem[{Buckley and Voorhees(2005)}]{buckley05}
Buckley C, Voorhees EM (2005) Retrieval system evaluation. In: Voorhees EM,
  Harman DK (eds) TREC: Experiment and Evaluation in Information Retrieval, The
  MIT Press, chap~2, pp 53--97

\bibitem[{Burges et~al(2005)Burges, Shaked, Renshaw, Lazier, Deeds, Hamilton,
  and Hullender}]{burges05}
Burges C, Shaked T, Renshaw E, Lazier A, Deeds M, Hamilton N, Hullender G
  (2005) Learning to rank using gradient descent. In: Proceedings of ICML 2005,
  pp 89--96

\bibitem[{Chapelle et~al(2009)Chapelle, Metzler, Zhang, and
  Grinspan}]{chapelle09}
Chapelle O, Metzler D, Zhang Y, Grinspan P (2009) Expected reciprocal rank for
  graded relevance. In: Proceedings of ACM CIKM 2009, pp 621--630

\bibitem[{Chapelle et~al(2011)Chapelle, Ji, Liao, Velipasaoglu, Lai, and
  Wu}]{chapelle11}
Chapelle O, Ji S, Liao C, Velipasaoglu E, Lai L, Wu SL (2011) Intent-based
  diversification of web search results: Metrics and algorithms. Information
  Retrieval 14(6):572--592

\bibitem[{Chen and Chen(2001)}]{chen01}
Chen KH, Chen HH (2001) The chinese text retrieval tasks of {NTCIR} workshop 2.
  In: Proceedings of NTCIR-2

\bibitem[{Chen et~al(2002)Chen, Chen, Kando, Kuriyama, Lee, Myaeng, Kishida,
  Eguchi, and Kim}]{chen02}
Chen KH, Chen HH, Kando N, Kuriyama K, Lee S, Myaeng SH, Kishida K, Eguchi K,
  Kim H (2002) Overview of {CLIR} task at the third {NTCIR} workshop. In:
  Proceedings of NTCIR-3

\bibitem[{Clarke et~al(2008)Clarke, Kolla, Cormack, Vechtomova, Ashkan,
  B\"{u}ttcher, and MacKinnon}]{clarke08}
Clarke CL, Kolla M, Cormack GV, Vechtomova O, Ashkan A, B\"{u}ttcher S,
  MacKinnon I (2008) Novelty and diversity in information retrieval evaluation.
  In: Proceedings of ACM SIGIR 2008, pp 659--666

\bibitem[{Cleverdon et~al(1966)Cleverdon, Mills, and Keen}]{cleverdon66}
Cleverdon CW, Mills J, Keen EM (1966) Factors determining the performance of
  indexing systems; volume 1: Design. Tech. rep., The College of Aeronautics,
  Cranfield

\bibitem[{Collins-Thompson et~al(2014)Collins-Thompson, Bennett, Diaz, Clarke,
  and Voorhees}]{collins-thompson14}
Collins-Thompson K, Bennett P, Diaz F, Clarke CL, Voorhees EM (2014) {TREC}
  2013 {Web Track} overview. In: Proceedings of TREC 2013

\bibitem[{Eguchi et~al(2003)Eguchi, Oyama, Ishida, Kando, and
  Kuriyama}]{eguchi03}
Eguchi K, Oyama K, Ishida E, Kando N, Kuriyama K (2003) Overview of the {Web
  Retrieval Task} at the third {NTCIR} workshop. In: Proceedings of NTCIR-3

\bibitem[{Eguchi et~al(2004)Eguchi, Oyma, Aizawa, and Ishikawa}]{eguchi04}
Eguchi K, Oyma K, Aizawa A, Ishikawa H (2004) Overview of the {Information
  Retrieval} task at {NTCIR}-4 {WEB}. In: Proceedings of NTCIR-4

\bibitem[{Fujii et~al(2004)Fujii, Iwayama, and Kando}]{fujii04}
Fujii A, Iwayama M, Kando N (2004) Overview of {Patent Retrieval} task at
  {NTCIR}-4. In: Proceedings of NTCIR-4

\bibitem[{Fujii et~al(2005)Fujii, Iwayama, and Kando}]{fujii05}
Fujii A, Iwayama M, Kando N (2005) Overview of {Patent Retrieval} task at
  {NTCIR}-5. In: Proceedings of NTCIR-5

\bibitem[{Fujii et~al(2007)Fujii, Iwayama, and Kando}]{fujii07}
Fujii A, Iwayama M, Kando N (2007) Overview of the {Patent Retrieval} task at
  the {NTCIR}-6 workshop. In: Proceedings of NTCIR-6, pp 359--365

\bibitem[{Gey et~al(2010)Gey, Larson, Kando, Machado, and Sakai}]{gey10}
Gey F, Larson R, Kando N, Machado J, Sakai T (2010) {NTCIR-GeoTime} overview:
  Evaluating geographic and temporal search. In: Proceedings of NTCIR-8, pp
  147--153

\bibitem[{Gey et~al(2011)Gey, Larson, Machado, and Yoshioka}]{gey11}
Gey F, Larson R, Machado J, Yoshioka M (2011) {NTCIR9-GeoTime} overview:
  Evaluating geographic and temporal search: Round 2. In: Proceedings of
  NTCIR-9, pp 9--17

\bibitem[{Golbus et~al(2013)Golbus, Aslam, and Clarke}]{golbus13}
Golbus PB, Aslam JA, Clarke CL (2013) Increasing evaluation sensitivity to
  diversity. Information Retrieval 16(4):530--555

\bibitem[{Harman(2005)}]{harman05}
Harman DK (2005) The {TREC} test collections. In: Voorhees EM, Harman DK (eds)
  TREC: Experiment and Evaluation in Information Retrieval, The MIT Press,
  chap~2, pp 21--52

\bibitem[{Hawking and Craswell(2005)}]{hawking05}
Hawking D, Craswell N (2005) The very large collection and web tracks. In:
  Voorhees EM, Harman DK (eds) TREC: Experiment and Evaluation in Information
  Retrieval, The MIT Press, chap~2, pp 200--231

\bibitem[{Hersh et~al(1994)Hersh, Buckley, Leone, and Hickam}]{hersh94}
Hersh W, Buckley C, Leone T, Hickam D (1994) {OHSUMED}: An interactive
  retrieval evaluation and new large test collection for research. In:
  Proceedings of ACM SIGIR 1994, pp 192--201

\bibitem[{Higashinaka et~al(2017)Higashinaka, Funakoshi, Inaba, Tsunomori,
  Takahashi, and Kaji}]{higashinaka17}
Higashinaka R, Funakoshi K, Inaba M, Tsunomori Y, Takahashi T, Kaji N (2017)
  Overview of {Dialogue Breakdown Detection Challenge} 3. In: Proceedings of
  Dialog System Technology Challenge 6 (DSTC6) Workshop

\bibitem[{Hu et~al(2015)Hu, Dou, Wang, and adn Ji-Rong~Wen}]{hu15}
Hu S, Dou Z, Wang X, adn Ji-Rong~Wen TS (2015) Search result diversification
  based on hierarchical intents. In: Proceedings of ACM CIKM 2015, pp 63--72

\bibitem[{Iwayama et~al(2003)Iwayama, Fujii, Kando, and Takano}]{iwayama03}
Iwayama M, Fujii A, Kando N, Takano A (2003) Overview of {Patent Retrieval}
  task at {NTCIR}-3. In: Proceedings of NTCIR-3

\bibitem[{J\"{a}rvelin and Kek\"{a}l\"{a}inen(2000)}]{jarvelin00}
J\"{a}rvelin K, Kek\"{a}l\"{a}inen J (2000) {IR} evaluation methods for
  retrieving highly relevant documents. In: Proceedings of ACM SIGIR 2000, pp
  41--48

\bibitem[{J\"{a}rvelin and Kek\"{a}l\"{a}inen(2002)}]{jarvelin02}
J\"{a}rvelin K, Kek\"{a}l\"{a}inen J (2002) Cumulated gain-based evaluation of
  {IR} techniques. ACM TOIS 20(4):422--446

\bibitem[{J\"{a}rvelin et~al(2008)J\"{a}rvelin, Price, Delcambre, and
  Nielsen}]{jarvelin08}
J\"{a}rvelin K, Price SL, Delcambre LM, Nielsen ML (2008) Discounted cumulated
  gain based evaluation of multiple-query {IR} sessions. In: Proceedings of
  ECIR 2008 (LNCS 4956), pp 4--15

\bibitem[{Joho et~al(2014)Joho, Jatowt, Blanco, Naka, and Yamamoto}]{joho14}
Joho H, Jatowt A, Blanco R, Naka H, Yamamoto S (2014) Overview of {NTCIR}-11
  {Temporal Information Access} ({Temporalia}) task. In: Proceedings of
  NTCIR-11, pp 429--437

\bibitem[{Joho et~al(2016)Joho, Jatowt, Blanco, Yu, and Yamamoto}]{joho16}
Joho H, Jatowt A, Blanco R, Yu H, Yamamoto S (2016) Overview of {NTCIR}-12
  {Temporal Information Access} ({Temporalia}-2) task. In: Proceedings of
  NTCIR-12, pp 217--224

\bibitem[{Kando et~al(1999)Kando, Kuriyama, Nozue, Eguchi, Kato, and
  Hidaka}]{kando99}
Kando N, Kuriyama K, Nozue T, Eguchi K, Kato H, Hidaka S (1999) Overview of
  {IR} tasks at the first {NTCIR} workshop. In: Proceedings of NTCIR-1, pp
  11--44

\bibitem[{Kando et~al(2001)Kando, Kuriyama, and Yoshioka}]{kando01}
Kando N, Kuriyama K, Yoshioka M (2001) Overview of {Japanese and English
  Information Retrieval} tasks (jeir) at the second {NTCIR} workshop. In:
  Proceedings of NTCIR-2

\bibitem[{Kato et~al(2017)Kato, Yamamoto, Manabe, Nishida, and Fujita}]{kato17}
Kato MP, Yamamoto T, Manabe T, Nishida A, Fujita S (2017) Overview of the
  {NTCIR}-13 {OpenLiveQ} task. In: Proceedings of NTCIR-13, pp 85--90

\bibitem[{Kato et~al(2018)Kato, Manabe, Fujita, Nishida, and Yamamoto}]{kato18}
Kato MP, Manabe T, Fujita S, Nishida A, Yamamoto T (2018) Challenges of
  multileaved comparison in practice: Lessons from {NTCIR}-13 {OpenLiveQ} task.
  In: Proceedings of ACM CIKM 2018, pp 1515--1518

\bibitem[{Kato et~al(2019)Kato, Nishida, Manabe, Fujita, and Yamamoto}]{kato19}
Kato MP, Nishida A, Manabe T, Fujita S, Yamamoto T (2019) Overview of the
  {NTCIR}-14 {OpenLiveQ}-2 task. In: Proceedings of NTCIR-14, p to appear

\bibitem[{Kishida(2005)}]{kishida05genap}
Kishida K (2005) Property of average precision and its generalization: An
  examination of evaluation indicator for information retrieval. Tech. Rep.
  NII-2005-014E, National Institute of Informatics

\bibitem[{Kishida et~al(2004)Kishida, Chen, Lee, Kuriyama, Kando, Chen, Myaeng,
  and Eguchi}]{kishida04}
Kishida K, Chen KH, Lee S, Kuriyama K, Kando N, Chen HH, Myaeng SH, Eguchi K
  (2004) Overview of {CLIR} task at the fourth {NTCIR} workshop. In:
  Proceedings of NTCIR-4

\bibitem[{Kishida et~al(2005)Kishida, Chen, Lee, Kuriyama, Kando, Chen, and
  Myaeng}]{kishida05}
Kishida K, Chen KH, Lee S, Kuriyama K, Kando N, Chen HH, Myaeng SH (2005)
  Overview of the {CLIR} task at the fifth {NTCIR} workshop (revised version).
  In: Proceedings of NTCIR-5

\bibitem[{Kishida et~al(2007)Kishida, Chen, Lee, Kuriyama, Kando, and
  Chen}]{kishida07}
Kishida K, Chen KH, Lee S, Kuriyama K, Kando N, Chen HH (2007) Overview of the
  {CLIR} task at the sixth {NTCIR} workshop. In: Proceedings of NTCIR-6, pp
  1--19

\bibitem[{Korfhage(1997)}]{korfhage97}
Korfhage RR (1997) Information Storage and Retrieval. John Wiley \& Sons

\bibitem[{Liu et~al(2014)Liu, Song, Zhang, Dou, Yamamoto, Kato, Ohshima, and
  Zhou}]{liu14}
Liu Y, Song R, Zhang M, Dou Z, Yamamoto T, Kato M, Ohshima H, Zhou K (2014)
  Overview of the {NTCIR}-11 {IMine} task. In: Proceedings of NTCIR-11, pp
  8--23

\bibitem[{Luo et~al(2017)Luo, Sakai, Liu, Dou, Xiong, and Xu}]{luo17}
Luo C, Sakai T, Liu Y, Dou Z, Xiong C, Xu J (2017) Overview of the {NTCIR}-13
  {We Want Web} task. In: Proceedings of NTCIR-13, pp 394--401

\bibitem[{Maddalena et~al(2017)Maddalena, Roitero, Demartini, and
  Mizzaro}]{maddalena17}
Maddalena E, Roitero K, Demartini G, Mizzaro S (2017) Considering assessor
  agreement in {IR} evaluation. In: Proceedings of ACM ICTIR 2017, pp 75--82

\bibitem[{Mao et~al(2019)Mao, Sakai, Luo, Xiao, Liu, and Dou}]{mao19}
Mao J, Sakai T, Luo C, Xiao P, Liu Y, Dou Z (2019) Overview of the {NTCIR}-14
  {We Want Web} task. In: Proceedings of NTCIR-14, p to appear

\bibitem[{Ohtsuka et~al(2004)Ohtsuka, Eguchi, and Yamana}]{ohtsuka04}
Ohtsuka T, Eguchi K, Yamana H (2004) An evaluation method of web search engines
  based on users' sense. In: Proceedings of NTCIR-4

\bibitem[{Oyama et~al(2004)Oyama, Eguchi, Ishikawa, and Aizawa}]{oyama04}
Oyama K, Eguchi K, Ishikawa H, Aizawa A (2004) Overview of the {NTCIR}-4 {WEB}
  {Navigational Retrieval} task 1. In: Proceedings of NTCIR-4

\bibitem[{Pollock(1968)}]{pollock68}
Pollock SM (1968) Measures for the comparison of information retrieval systems.
  American Documentation 19(4):387--397

\bibitem[{Robertson and Hull(2001)}]{robertson01}
Robertson S, Hull DA (2001) The {TREC}-9 filtering track final report. In:
  Proceedings of TREC-9

\bibitem[{Robertson et~al(2010)Robertson, Kanoulas, and Yilmaz}]{robertson10}
Robertson SE, Kanoulas E, Yilmaz E (2010) Extending average precision to graded
  relevance judgments. In: Proceedings of ACM SIGIR 2010, pp 603--610

\bibitem[{Sakai(2004)}]{sakai04ntcir}
Sakai T (2004) New performance metrics based on multigrade relevance: Their
  application to question answering. In: Proceedings of NTCIR-4 (Open
  Submission Session)

\bibitem[{Sakai(2006{\natexlab{a}})}]{sakai06sigir}
Sakai T (2006{\natexlab{a}}) Evaluating evaluation metrics based on the
  bootstrap. In: Proceedings of ACM SIGIR 2006, pp 525--532

\bibitem[{Sakai(2006{\natexlab{b}})}]{sakai06ipsj}
Sakai T (2006{\natexlab{b}}) For building better retrieval systems: Trends in
  information retrieval evaluation based on graded relevance (in japanese).
  IPSJ Magazine 47(2):147--158

\bibitem[{Sakai(2007{\natexlab{a}})}]{sakai07sigir}
Sakai T (2007{\natexlab{a}}) Alternatives to bpref. In: Proceedings of ACM
  SIGIR 2007, pp 71--78

\bibitem[{Sakai(2007{\natexlab{b}})}]{sakai07evia}
Sakai T (2007{\natexlab{b}}) On penalising late arrival of relevant documents
  in information retrieval evaluation with graded relevance. In: Proceedings of
  EVIA 2007, pp 32--43

\bibitem[{Sakai(2007{\natexlab{c}})}]{sakai07tod}
Sakai T (2007{\natexlab{c}}) On the properties of evaluation metrics for
  finding one highly relevant document. IPSJ Digital Courier 3:643--660,
  \urlprefix\url{https://doi.org/10.2197/ipsjdc.3.643}

\bibitem[{Sakai(2007{\natexlab{d}})}]{sakai07ipm}
Sakai T (2007{\natexlab{d}}) On the reliability of information retrieval
  metrics based on graded relevance. Information Processing and Management
  43(2):531--548

\bibitem[{Sakai(2012)}]{sakai12www}
Sakai T (2012) Evaluation with informational and navigational intents. In:
  Proceedings of WWW 2012, pp 499--508

\bibitem[{Sakai(2014)}]{sakai14promise}
Sakai T (2014) Metrics, statistics, tests. In: PROMISE Winter School 2013:
  Bridging between Information Retrieval and Databases (LNCS 8173), pp 116--163

\bibitem[{Sakai(2017)}]{sakai17evia-unan}
Sakai T (2017) Unanimity-aware gain for highly subjective assessments. In:
  Proceedings of EVIA 2017, pp 39--42

\bibitem[{Sakai(2018)}]{sakai18short}
Sakai T (2018) Comparing two binned probability distributions for information
  access evaluation. In: Proceedings of ACM SIGIR 2018, pp 1073--1076

\bibitem[{Sakai and Kando(2008)}]{sakai08irj}
Sakai T, Kando N (2008) On information retrieval metrics designed for
  evaluation with incomplete relevance assessments. Information Retrieval
  11:447--470

\bibitem[{Sakai and Robertson(2008)}]{sakai08evia}
Sakai T, Robertson S (2008) Modelling a user population for designing
  information retrieval metrics. In: Proceedings of EVIA 2018, pp 30--41

\bibitem[{Sakai and Song(2011)}]{sakai11sigir}
Sakai T, Song R (2011) Evaluating diversified search results using per-intent
  graded relevance. In: Proceedings of ACM SIGIR 2011, pp 1043--1052

\bibitem[{Sakai and Song(2013)}]{sakai13irj}
Sakai T, Song R (2013) Diversified search evaluation: Lessons from the
  {NTCIR}-9 {INTENT} task. Information Retrieval 16(4):504--529

\bibitem[{Sakai et~al(1999)Sakai, Kitani, Ogawa, Ishikawa, Kimoto, Keshi,
  Toyoura, Fukushima, Matsui, Ueda, Tokunaga, Tsuruoka, Nakawatase, Agata, and
  Kando}]{sakai99bmir}
Sakai T, Kitani T, Ogawa Y, Ishikawa T, Kimoto H, Keshi I, Toyoura J, Fukushima
  T, Matsui K, Ueda Y, Tokunaga T, Tsuruoka H, Nakawatase H, Agata T, Kando N
  (1999) {BMIR-J2}: A test collection for evaluation of japanese information
  retrieval systems. 33 (1):13--17

\bibitem[{Sakai et~al(2008)Sakai, Kando, Lin, Mitamura, Shima, Ji, Chen, and
  Nyberg}]{sakai08ir4qa}
Sakai T, Kando N, Lin CJ, Mitamura T, Shima H, Ji D, Chen KH, Nyberg E (2008)
  Overview of the {NTCIR}-7 {ACLIA} {IR4QA} task. In: Proceedings of NTCIR-7,
  pp 77--114

\bibitem[{Sakai et~al(2010{\natexlab{a}})Sakai, Ishikawa, and
  Kando}]{sakai10cqa}
Sakai T, Ishikawa D, Kando N (2010{\natexlab{a}}) Overview of the {NTCIR}-8
  {Community QA} pilot task (part {II}): System evaluation. In: Proceedings of
  NTCIR-8, pp 433--457

\bibitem[{Sakai et~al(2010{\natexlab{b}})Sakai, Shima, Kando, Song, Lin,
  Mitamura, Sugimoto, and Lee}]{sakai10ir4qa}
Sakai T, Shima H, Kando N, Song R, Lin CJ, Mitamura T, Sugimoto M, Lee CW
  (2010{\natexlab{b}}) Overview of {NTCIR}-8 {ACLIA} {IR4QA}. In: Proceedings
  of NTCIR-8, pp 63--93

\bibitem[{Sakai et~al(2011)Sakai, Ishikawa, Kando, Seki, Kuriyama, and
  Lin}]{sakai11wsdm}
Sakai T, Ishikawa D, Kando N, Seki Y, Kuriyama K, Lin CY (2011) Using
  graded-relevance metrics for evaluating community {QA} answer selection. In:
  Proceedings of ACM WSDM 2011, pp 187--196

\bibitem[{Sakai et~al(2013)Sakai, Dou, Yamamoto, Liu, Zhang, and
  Song}]{sakai13intent}
Sakai T, Dou Z, Yamamoto T, Liu Y, Zhang M, Song R (2013) Overview of the
  {NTCIR}-10 {INTENT}-2 task. In: Proceedings of NTCIR-10, pp 94--123

\bibitem[{Sakai et~al(2019)Sakai, Ferro, Soboroff, Zeng, Xiao, and
  Maistro}]{sakai19centre}
Sakai T, Ferro N, Soboroff I, Zeng Z, Xiao P, Maistro M (2019) Overview of the
  {NTCIR}-14 {CENTRE} task. In: Proceedings of NTCIR-14, p to appear

\bibitem[{Shang et~al(2016)Shang, Sakai, Lu, Li, Higashinaka, and
  Miyao}]{shang16}
Shang L, Sakai T, Lu Z, Li H, Higashinaka R, Miyao Y (2016) Overview of the
  {NTCIR}-12 {Short Text Conversation} task. In: Proceedings of NTCIR-12, pp
  473--484

\bibitem[{Shang et~al(2017)Shang, Sakai, Li, Higashinaka, Miyao, Arase, and
  Nomoto}]{shang17}
Shang L, Sakai T, Li H, Higashinaka R, Miyao Y, Arase Y, Nomoto M (2017)
  Overview of the {NTCIR}-13 {Short Text Conversation} task. In: Proceedings of
  NTCIR-13, pp 194--210

\bibitem[{Song et~al(2011)Song, Zhang, Sakai, Kato, Liu, Sugimoto, Wang, and
  Orii}]{song11}
Song R, Zhang M, Sakai T, Kato MP, Liu Y, Sugimoto M, Wang Q, Orii N (2011)
  Overview of the {NTCIR}-9 {INTENT} task. In: Proceedings of NTCIR-9, pp
  82--105

\bibitem[{Sormunen(2002)}]{sormunen02}
Sormunen E (2002) Liberal relevance criteria of {TREC}: Counting on negligible
  documents? In: Proceedings of ACM SIGIR 2002, pp 324--330

\bibitem[{Tang et~al(2013)Tang, Kang, Kimura, Lee, and Trotman}]{tang13}
Tang LX, Kang IS, Kimura F, Lee YH, Trotman A (2013) Overview of the {NTCIR}-10
  {Cross-lingual Link Discovery} task. In: Proceedings of NTCIR-10, pp 8--38

\bibitem[{Voorhees(2006)}]{voorhees06}
Voorhees EM (2006) Ellen m. voorhees. In: Overview of the {TREC} 2005 {Robust
  Retrieval} Track

\bibitem[{Wang et~al(2018)Wang, Wen, Dou, Sakai, and Zhang}]{wang18}
Wang X, Wen JR, Dou Z, Sakai T, Zhang R (2018) Search result diversity
  evaluation based on intent hierarchies. IEEE TKDE 30(1):156--169

\bibitem[{Yamamoto et~al(2016)Yamamoto, Liu, Zhang, Dou, Zhou, Markov, Kato,
  Ohshima, and Fujita}]{yamamoto16}
Yamamoto T, Liu Y, Zhang M, Dou Z, Zhou K, Markov I, Kato MP, Ohshima H, Fujita
  S (2016) Overview of the {NTCIR}-12 {IMine}-2 task. In: Proceedings of
  NTCIR-12, pp 8--26

\bibitem[{Yasukawa et~al(2014)Yasukawa, Diaz, Druck, and Tsukada}]{yasukawa14}
Yasukawa M, Diaz F, Druck G, Tsukada N (2014) Overview of the {NTCIR}-11
  {Cooking Recipe Search} task. In: Proceedings of NTCIR-11, pp 483--496

\bibitem[{Zanibbi et~al(2016)Zanibbi, Aizawa, Kohlhase, Ounis, Topi\'{c}, and
  Davila}]{zanibbi16}
Zanibbi R, Aizawa A, Kohlhase M, Ounis I, Topi\'{c} G, Davila K (2016)
  {NTCIR}-12 {MathIR} task overview. In: Proceedings of NTCIR-12, pp 299--308

\bibitem[{Zhai et~al(2003)Zhai, Cohen, and Lafferty}]{zhai03}
Zhai C, Cohen WW, Lafferty J (2003) Beyond independent relevance: Methods and
  evaluation metrics for subtopic retrieval. In: Proceedings of ACM SIGIR 2003,
  pp 10--17

\bibitem[{Zhang and Huang(2019)}]{zhang19}
Zhang Y, Huang M (2019) Overview of the {NTCIR}-14 {Short Text Generation}
  subtask: {Emotion Generation Challenge}. In: Proceedings of NTCIR-14, p to
  appear

\bibitem[{Zhou et~al(2013)Zhou, Lalmas, Sakai, Cummins, and Jose}]{zhou13}
Zhou K, Lalmas M, Sakai T, Cummins R, Jose JM (2013) On the reliability and
  intuitiveness of aggregated search metrics. In: Proceedings of ACM CIKM 2013,
  pp 689--698

\end{thebibliography}
